\documentclass[twocolumn]{aastex701}
\usepackage{amssymb, amsmath, amsbsy, booktabs}
\usepackage{subcaption}
\usepackage{CJKutf8}
\newcommand{\mgal}{\ensuremath{M_\star}}

\newcommand{\lbol}{\ensuremath{L\mathrm{_{bol}}}}

\newcommand{\ledd}{\ensuremath{L\mathrm{_{Edd}}}}
\newcommand{\redd}{\ensuremath{L_\mathrm{bol}/L_{\mathrm{Edd}}}}

\newcommand{\lfive}{\ensuremath{\lambda L_{\lambda}(5100~\AA)}}

\newcommand{\msun}{\ensuremath{M_{\odot}}}

\newcommand{\ergs}{\ensuremath{\mathrm{erg~s^{-1}}}}
\newcommand{\kms}{\ensuremath{\mathrm{km~s^{-1}}}}
\newcommand{\mbh}{\ensuremath{M_\mathrm{BH}}}

\newcommand{\chisq}{\ensuremath{\chi^2}}

\newcommand{\ha}{\ensuremath{\mathrm{H\alpha}}}
\newcommand{\hb}{\ensuremath{\mathrm{H\beta}}}

\newcommand{\oiii}{[O\,{\footnotesize III}]}

\newcommand{\feii}{{\rm Fe\,{\footnotesize II}}}

\newcommand{\mgii}{Mg\,{\footnotesize II}}

\newcommand{\neiii}{[Ne\,{\footnotesize III}] }
\newcommand{\nev}{[Ne\,{\footnotesize V}]}
\newcommand{\oii}{[O\,{\footnotesize II}]}
\newcommand{\oi}{[O\,{\footnotesize I}]}

\begin{document}
\begin{CJK*}{UTF8}{gbsn}
\title{A New Sample of $\sim$ 100 Intermediate-mass Black Holes Reaching $z \approx 1$
}

\author{Wen-Juan Liu (刘文娟)}
\affiliation{Yunnan Observatories, Chinese Academy of Sciences, 
Kunming, Yunnan 650011, China}
\affiliation{Key Laboratory for Polar Science, Ministry of Natural Resources, 
Polar Research Institute of China, Shanghai 200136, China}
\email[show]{wjliu@ynao.ac.cn}  

\author{Luis C. Ho}
\affiliation{Kavli Institute for Astronomy and Astrophysics, Peking University, 
Beijing 100871, China}
\affiliation{Department of Astronomy, School of Physics, Peking University, 
Beijing 100871, China}
\email{lho.pku@gmail.com}

\author{Su Yao (姚苏)}
\affiliation{National Astronomical Observatories, Chinese Academy of Sciences,
Beijing 100101, China}
\email{yaosu@bao.ac.cn}

\author{Xiao-Bo Dong (董小波)}
\affiliation{Yunnan Observatories, Chinese Academy of Sciences, 
Kunming, Yunnan 650011, China}
\email{xbdong@ynao.ac.cn}

\author{Yaqi Zhao (赵雅琪)}
\affiliation{Department of Astronomy, University of Science and Technology of China, Hefei 230026, China}
\affiliation{Polar Research Institute of China, 451 Jinqiao Road, Pudong, Shanghai 200136, Peopleʼs Republic of China}
\affiliation{Key Laboratory for Polar Science, Ministry of Natural Resources, 
Polar Research Institute of China, Shanghai 200136, China}
\email{zxmyg86400@mail.ustc.edu.cn}

\begin{abstract}

We present a systematic search for intermediate-mass black hole (IMBH) active galactic nuclei (AGNs)
at $0.5<z\lesssim1$ using DESI DR1 spectroscopy.
We identify 98 broad-line IMBH AGNs with black hole masses $\mbh<10^{6}$ \msun\ through quantitative 
spectral decomposition and broad-\hb\ selection.
This sample spans \mbh=$10^{5.5} -10^{6.0}$ \msun\ and Eddington ratios from 1.2 to 9.0
extending systematic IMBH AGN searches to intermediate redshift.
Compared with a consistently selected $z<0.6$ IMBH sample, 
the $0.5<z<1$ sources reach higher broad-\hb\ luminosities and substantially higher Eddington ratios, 
indicating more extreme accretion states at earlier cosmic times.
They also show broader \oiii\ profiles and stronger blueshifted wing components,
with these kinematic differences persisting after matching in black hole mass and Eddington ratio.
These results reveal two distinct signatures of evolution among IMBH AGNs at $z<1$:
ability of IMBHs to reach increasingly extreme accretion states toward higher redshift, 
and systematic changes in their ionized-gas kinematics.
The former demonstrates that rapid, including super-Eddington, 
growth of IMBHs can persist to relatively late cosmic times, 
while the latter may indicate evolution in the ionized-gas environment 
and associated outflow activity of actively growing IMBHs.
Together, these findings provide new constraints on the evolutionary pathways of IMBHs at $z<1$, 
and show that seed-mass black holes can continue to undergo rapid growth well after the cosmic dawn.
\end{abstract}



\section{Introduction} 
Intermediate-mass black holes (IMBHs; $10^2$--$10^6$ \msun),   
occupy the sparsely populated region of the black hole (BH) mass spectrum,
bridging stellar-mass BHs and supermassive black holes (SMBHs) 
that reside in the galaxy centers.
This intermediate mass regime is of special importance because SMBHs are widely hypothesized 
to originate from ``seed'' BHs formed in the early Universe through mechanisms such as 
remnants of Population III stars or the direct collapse of gas clouds \citep{volonteri2012,Greene2020,Inayoshi2020,volonteri2021}.
Yet neither the seeds themselves nor their subsequent growth pathways have been directly observed,
leaving the connection between the hypothetical seeds and SMBHs seen across cosmic time poorly constrained.
IMBHs therefore offer a uniquely accessible observational handle on this missing evolutionary phase of SMBHs,
and their demographics offer critical tests for SMBH seeding models, accretion process, and BH-galaxy co-evolution. 

Observational studies of IMBHs and IMBH AGNs have so far been concentrated primarily in the local Universe at $z\lesssim0.5$,
while recent JWST observations have begun to probe the low-mass end of the BH population at much higher redshifts ($z>4$).
Some of the earliest and most compelling low-mass BH candidates in the local Universe were identified 
in the nearby galaxies NGC 4395 and POX 52 \citep{Filippenko2003,Barth2004}.
More recently, hundreds of IMBH candidates have been reported through optical broad-line AGN signatures \citep[e.g.,][]{gh07,imbh_dong12,imbh_liu18,Pucha2025,imbh_liu26}, 
X-ray and multiwavelength observations selection \citep[e.g.,][]{Desroches2009,Lemons2015,She2017,Mezcua2018,Bi2020,Reines2020,Eberhard2025}, 
optical/UV variability \citep[e.g.,][]{Baldassare2018,Jorge_2020_imbh_vari,Ward_2022_imbh_vari}, 
tidal disruption events \citep[e.g.,][]{Jin_imbh_tde_2025,Wangjialai_2025_tde}, 
and a handful of dynamical detections \citep[e.g.,][]{Haberle2024,Huang2025} . 
These studies have significantly advanced our understanding of IMBH across multiple fronts, 
spanning their statistical demographics, multiwavelength properties, accretion characteristics,  
host-galaxy properties, and the BH--galaxy scaling relations (\mbh-\mgal, \mbh-$\sigma_{\star}$)\citep[e.g.,][]{Kormendy2013,Greene2020},
At the other extreme, recent observations with James Webb Space Telescope (JWST) 
have revealed numerous low-mass AGNs at $z=4-10$ \citep[e.g.,][]{GNZ11_2024,Geris2026}, 
approaching the IMBH mass regime.
JWST has also uncovered a population of ``little red dots'' (LRDs), 
many of which appear to host low-mass or moderately accreting black holes \citep{Kocevski2023,Matthee2024_lrd,Akins2025,chenchanghao2025,Inayoshi2025}.
These discoveries suggest accelerated BH growth in early dense environments 
and reveal systems that appear ``overmassive'' relative to their hosts compared to local scaling relations.

However, the vast intermediate redshift range of $z\approx 0.5-4$ remains sparsely explored for IMBHs,
despite being essential for linking the well-characterized local IMBH population 
to the newly emerging high-redshift low-mass AGN population.
The primary difficulty is that IMBH AGNs are intrinsically faint,
which poses detection challenges even in the local Universe.
At intermediate redshifts, this difficulty becomes far more severe. 
Existing wide-area spectroscopic surveys - most notably SDSS - 
do not reach the depth required to identify such faint systems efficiently. 
Although JWST has sufficient ability to probe IMBHs in this regime, 
its small field of view, pointed observing strategy, 
and limited allocation time have naturally concentrated efforts on the highest-redshift Universe, 
leading to few systematic searches at intermediate redshifts.

As a result, the cosmic evolution of IMBHs from high-$z$ seeds to present-day descendants remains poorly constrained.
This redshift interval is astrophysically important because it encompasses the period of ``cosmic noon'', 
the peak epoch of galaxy and SMBH growth.
Cosmic noon is marked by elevated star formation rates, abundant cold-gas supply, 
and vigorous SMBH accretion activity across the cosmic time.
It is a critical turning point in galaxy and AGN evolution:
beyond this epoch, both star formation and SMBH accretion begin their long-term decline toward the present day.
Exploring IMBH in this ``middle-age'' Universe is therefore essential for tracing when and how 
low-mass end of SMBHs evolve within the broader context of cosmic baryon cycling, 
and for linking the well-characterized IMBH populations in the local Universe to the rapidly growing BHs observed at higher redshifts, 
thereby building a continuous picture of BH demographics through cosmic time.

In this paper, we address this gap by conducting the first systematic search for IMBH AGN candidates at $z\approx0.5-1$
using spectroscopic data from the Dark Energy Spectroscopic Instrument (DESI) Survey Data Release 1 (DR1) \citep{DESIDR1}.
DESI Survey combines large sky coverage, higher spectral resolution ($R\sim4000$), 
and substantially greater depth than previous wide-area surveys (e.g., SDSS),
enabling the detection of much fainter AGN populations. 
In addition, DESI's wavelength coverage simultaneously includes \hb\ and \mgii\ at $z\approx 0.5-1$,
providing the key broad-line diagnostics required for robust IMBH AGN identification.
Taking advantage of these capabilities, we identified 98 broad-line IMBH AGN candidates with \mbh\ $\leqslant10^6$ \msun,
significantly extending IMBH searches into this previously underexplored redshift regime.

This paper is structured as follows: 
Section~2 describes the spectroscopic dataset used in this work.
Section~3 details the spectral fitting methodology and the construction of the IMBH AGN sample.
In Section~4, we present the sample properties, investigate the accretion-related spectral sequence commonly referred to as Eigenvector~1 (EV1), and examine the kinematics of the \oiii\ emission.
Section~5 discusses the implications of our results in the context of IMBH growth and evolution.
Section~6 summarizes our main conclusions.
Throughout this paper, we adopt a cosmology with $H_{0}=70$\,\kms\,Mpc$^{-1}$, $\Omega_\mathrm{m}=0.3$, and $\Omega_{\Lambda}=0.7$.

\section{Spectroscopy Data and Spectral Analysis}

\subsection{Spectroscopy Data}

The IMBH AGN candidates analyzed in this work are drawn from the DESI Survey DR1 \citet{DESIDR1}.
DESI is a massively multiplexed fiber-based spectroscopic survey operating 
on the 4-meter Mayall Telescope at Kitt Peak National Observatory. 
Its 5000 robotically positioned fibers enable simultaneous observations 
across a 3.2$\arcdeg$-diameter field of view, 
providing unprecedented survey speed and sky coverage.
DESI spectra span a wavelength range of 3600--9800 \AA\ and are obtained through 
three spectrograph channels: blue (3600--5800 \AA), red (5600--7600 \AA), and near-infrared (7400--9800 \AA), 
with overlapping bandpasses that ensure continuous spectral coverage.
Each channel has a resolving power of $R \sim 2000-5500$,
corresponding to a substantial improvement over previous wide-area surveys such as SDSS.

DESI DR1 contains data from the first 13 months of the main survey (May 2021--June 2022),
covering over 9000 deg$^2$ of sky and providing reliable redshifts for 18.7 million unique targets. 
These observations comprise DESI's major extragalactic targets classes:
Bright Galaxy Survey (BGS, $0<z<0.6$), Luminous Red Galaxies (LRGs, $0.4<z<1.1$), 
Emission-Line Galaxies (ELGs, $0.6<z<1.6$), and QSOs ($0.9<z<4$).
Together, they provide a comprehensive census of galaxies and AGNs across $0<z<4$.
The DESI DR1 spectra are processed using the DESI spectroscopic pipeline \citep{DESIspecpipeline},
and redshifts are obtained with the Redrock \citep{DESIredshift}.
The release provides both per-exposure spectra and survey-level coadds,
with the latter referred to as healpix-coadd-individual products that
combine multiple exposures of the same target obtained within a single DESI survey. 
These spectra serve as the baseline spectral products used in this work.

To construct our parent sample,
we adopt the DR1-recommended requirement ZWARN $=0$,
which identifies spectra with no known issues in either the input data or redshift fit.
We then select objects classified as GALAXY or QSO and restrict range to $0.5\leq z \leq 0.956$.
The lower boundary at $z=0.5$ ensures that \ha\ is redshifted out of 
the DESI wavelength coverage, allowing this study to focus exclusively 
on \hb\ and \mgii\ for broad-line IMBH identification; 
a complementary \ha-based search will be presented in a future work.
The upper boundary at $z<0.956$ guarantees that \oiii$\lambda5007$,
a key diagnostic feature for line-width modeling and AGN verification,
remains within the DESI spectral range.
This selection yields a parent sample of 4026945 sources.
For targets with multiple available spectra in DR1, 
arising from repeated observations from different surveys, 
we select the highest signal-to-noise (S/N) spectrum for spectral fitting. 
All spectra are corrected for Galactic extinction using the dust map of \citet{Schlegel1998} and the reddening curve of \citet{Fitzpatrick1999}.
Each spectrum is then shifted to the rest frame using the redshifts provided by the SDSS pipeline.

\subsection{Spectral Fitting}

To identify IMBH AGN candidates and measure their BH masses, 
we perform detailed spectral fitting in two regions centered on \hb\ and \mgii, respectively.
Owing to the redshift range of our sample ($0.5\leqslant z\lesssim 1$),
the \ha\ region is not covered by the DESI spectra.
Therefore, the identification of broad-line emission in our sample 
is anchored primarily on the detection of a broad \hb\ component,
with the presence of broad \mgii\ emission serving as a supporting and consistency check where available.
The decomposition of broad and narrow components is primarily constrained by the narrow \hb\ line profile itself and by the kinematics of the narrow \oiii$\lambda\lambda4959,5007$ lines,
which together provide the main empirical reference for the narrow-line component.
When available, additional narrow lines such as \oii$\lambda3727$ and \oi$\lambda6300$ are also used as supplementary constraints to assess the consistency of the narrow-line kinematics.

\subsubsection{\hb+\oiii\ Region}

The \hb\ region provides the primary diagnostic for identifying broad-line AGNs in our sample.
This region includes \hb\ emission line, the \oiii$\lambda\lambda4959,5007$ doublet, 
as well as AGN power-law emission, host-galaxy starlight, and optical \feii\ blends.

We first model the continuum over the wavelength range 3700-5500 \AA,
which covers the \hb\ region and the adjacent optical features used in this work.
This wavelength interval allows reliable modeling of host-galaxy stellar continuum 
and the optical \feii\ emission. 
Owing to the fiber aperture (1.5\arcsec) of DESI spectroscopy as well as the
intrinsically weak nuclear emission of IMBH AGNs,
the observed optical spectra often contain non-negligible contributions from host-galaxy starlight.
At $0.5<z<1$, this effect is further enhanced because a fixed fiber aperture corresponds to a larger physical scale, 
increasing the contribution of host-galaxy light within the observed spectrum.
Accurate modeling and subtraction of the stellar continuum in the \hb\ region is therefore
essential for reliable measurements of both broad and narrow emission-line components.

The continuum modeling follows the procedure established in our previous IMBH AGN 
studies \citep{imbh_dong12,imbh_liu18,imbh_liu26}, and is briefly summarized here.
The observed continuum is represented as a linear combination of three components:
host-galaxy starlight, AGN nuclear continuum, and optical \feii\ multiplets. 
The starlight component is modeled using six synthetic galaxy templates constructed by 
\citet{lu2006} from the \citet{bc03} stellar population model.
The templates are convolved with a Gaussian kernel to account for stellar velocity broadening and instrumental resolution.
Because the stellar absorption features are weak or unresolved in most spectra,
the stellar velocity broadening cannot be reliably measured from the data alone.
We therefore restrict the template broadening to 30 to 150 \kms, 
a physically plausible range broadly expected for low-mass AGN host galaxies based 
on the low-mass extrapolation of the $\mbh-\sigma_{\star}$.
The fitted broadening is used only as a nuisance parameter in the continuum 
decomposition and is not interpreted as a direct measurement of $\sigma_{\star}$.

convolved with a Gaussian kernel to match the widths of stellar absorption features.
Small velocity shifts relative to the catalog redshift are allowed during fitting.
The AGN continuum is modeled as a single power law.
The optical \feii\ emission is modeled using analytical templates 
of \citet{dong11}, following our previous low-redshift IMBH AGN studies \citep{imbh_dong12,imbh_liu18,imbh_liu26}.
The templates are constructed from the well-resolved \feii\ multiplets of I~Zw~1 \citep{veron04} 
and include both broad and narrow \feii\ components.
They adopt the relative wavelengths and intensities measured in I~Zw~1, while the normalization, velocity width, and velocity shift of each component are fitted for individual objects.
The best-fitting continuum model is determined by minimizing $\chisq$.
In practice, continuum modeling and emission-line fitting are performed in an iterative manner, 
following the strategy adopted in our previous IMBH AGN studies \citep{imbh_dong12,imbh_liu18,imbh_liu26}.
Since reliable emission-line measurements require accurate continuum subtraction, 
while continuum fitting itself depends on proper masking of emission lines, 
these two components are refined jointly through successive iterations until convergence is achieved.

We model the emission lines in the \hb+\oiii\ region,
including both the broad and narrow components of \hb\ and the \oiii$\lambda\lambda4959,5007$ doublet.
The broad \hb\ component is modeled using one or two Gaussians.
Given that broad emission lines in IMBH AGNs are typically relatively weak and narrow,
a single Gaussian is adopted as the default model, 
with a second Gaussian introduced only when statistically justified by an $F-$test ($P<0.05$).
Both Gaussian components are required to share the same kinematic parameters across the doublet, 
and the adoption of a two-component model is justified only when it yields 
a statistically significant improvement according to an $F$-test ($P < 0.05$).

\begin{figure*}[htbp]
   \centering
   \begin{subfigure}{0.9\textwidth}
       \centering
       \includegraphics[width=\textwidth]{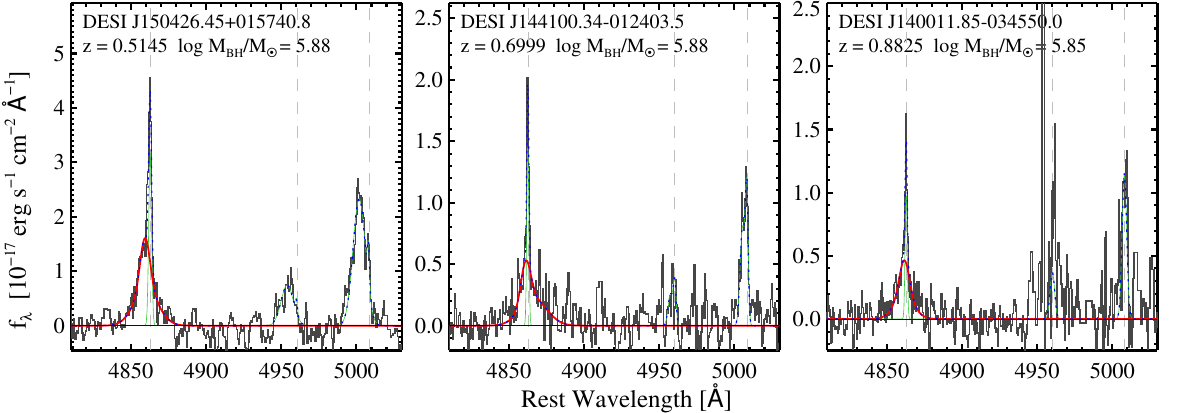}
       \caption{}
   \end{subfigure}
   \begin{subfigure}{0.9\textwidth}
       \centering
       \includegraphics[width=\textwidth]{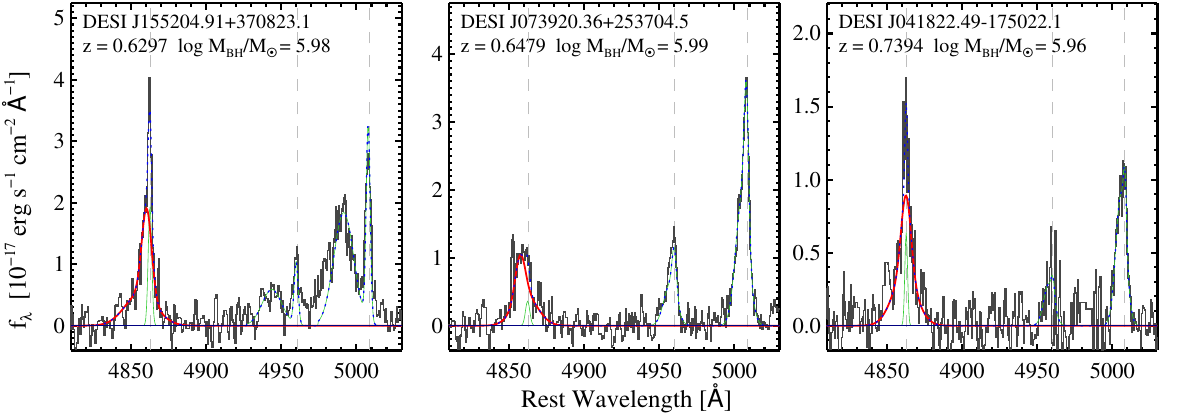}
       \caption{}
   \end{subfigure}
   \begin{subfigure}{0.9\textwidth}
       \centering
       \includegraphics[width=\textwidth]{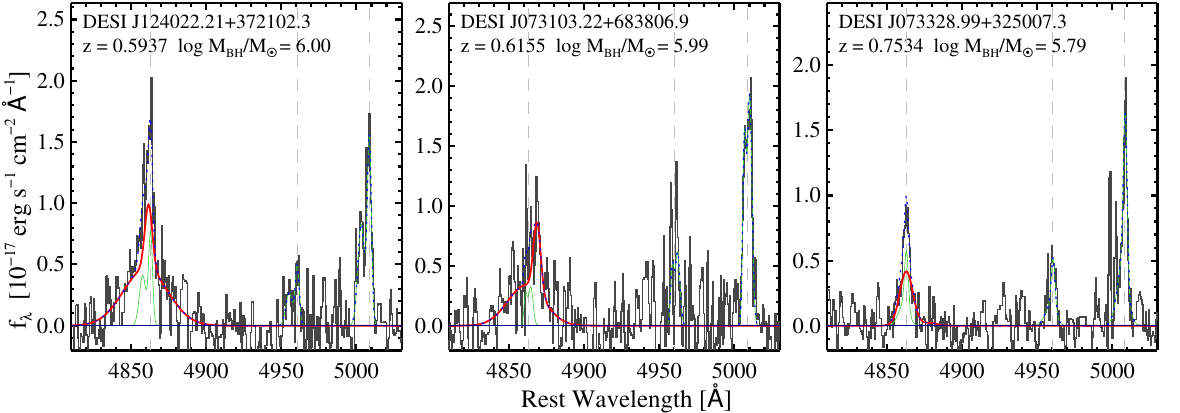}
       \caption{}
   \end{subfigure}
   \caption{Illustration of the fitting schemes adopted for the narrow \hb\ component in the \hb+\oiii\ region. 
   All panels show the continuum-subtracted emission-line spectra over 4800--5050~\AA, 
   with the observed spectra in black, the broad and narrow components in red and green, respectively. 
   (a) Examples in which the narrow \hb\ component can be well separated from the broad \hb\ emission and is fitted independently. 
   (b) Examples in which the \oiii\ profile requires an additional wing component, 
   but the \oiii\ wing is not an appropriate template for narrow \hb; in these cases, 
   narrow \hb\ is tied only to the core component of \oiii. 
   (c) Examples in which narrow \hb\ is tied to the full \oiii\ profile, including all \oiii\ kinematic components. 
   These schemes are selected on an object-by-object basis according to the 
   separability of narrow \hb\ from the broad component and the detailed \oiii\ line profile.
   }\label{fig:fig1}
\end{figure*}

To robustly separate the narrow and broad components of \hb, 
we adopt a classification-based approach motivated by the diversity of observed line profiles.
A common approach is to use a strong narrow emission line, such as \oiii, 
as an empirical reference to constrain the narrow component of Balmer lines.
In practice, however, the degree to which of the \oiii\ profile can be directly applied varies from source to source.
Based on detailed inspection of the spectra, 
we find that the correspondence between \hb\ and \oiii\ is not uniform across the sample. 
In some sources, the narrow \hb\ component is consistent with the core of \oiii, 
while in others it can be constrained directly from its own profile or remains only weakly constrained due to limited S/N.
We identify three representative regimes based on the observed line morphology (see Figure~\ref{fig:fig1}).
(1) In sources where the narrow and broad components of \hb\ are clearly distinguishable from the line profile itself,
the decomposition can be performed directly without reference to \oiii.
These cases provide the most robust constraints on the broad \hb\ properties.
(2) In sources where \oiii\ exhibits strongly asymmetric profiles,
often with pronounced blueshifted wings or even a globally blueshifted profile,
while \hb\ remains relatively symmetric and centered near the systemic velocity, 
the full \oiii\ profile does not provide a good template for narrow \hb.
In such cases, the narrow \hb\ component is constrained using only the core component of \oiii.
(3) In sources where the \oiii\ profile shows significant asymmetry or extended structure, 
while the \hb\ profile does not provide sufficient independent constraints on the narrow-line shape, 
the narrow \hb\ component is constrained using the full \oiii\ profile. 
This object-by-object treatment reflects the diversity of \oiii\ profiles and the limited 
ability of the \hb\ line alone to constrain the narrow component in some spectra. 
Although outflowing gas in the narrow-line region may emit both forbidden and recombination 
lines, its observational signatures can differ between \oiii\ and \hb\ because of differences 
in emissivity and ionization structure. 
In addition, the limited S/N of our spectra and the presence of broad \hb\ emission further 
reduce the detectability of weak asymmetric features in the narrow \hb\ component.
The frequent presence of asymmetric and blueshifted components in the \oiii\ profiles also 
motivates the systematic investigation of their kinematic properties presented in \S~\ref{subsec:o3profile}.

All three schemes are explored for each source to assess the range of plausible decompositions of the \hb\ profile.
This procedure allows us to evaluate whether the inferred broad-\hb\ component is robust 
against different plausible treatments of the narrow \hb\ emission.
Rather than relying solely on the minimum \chisq, 
the selection of the adopted model is guided by a combination of statistical goodness-of-fit,
consistency with the observed line profiles, and physical plausibility.
In particular, preference is given to solutions that provide a coherent description of both \hb\ and \oiii, 
without introducing spurious broad components or overfitting asymmetric narrow-line structures.
The resulting decomposition provides the basis for the subsequent assessment of the presence and reliability of broad \hb\ emission, 
as described in \S~\ref{subsec:creteria}.

Finally, we also fit the \oii$\lambda\lambda3726,3728$ doublet independently 
using a procedure analogous to \oiii.
Each line is modeled with one or two Gaussians as needed, 
with the wavelength separation fixed to the laboratory value and 
kinematic parameters constrained consistently across the doublet.

\begin{figure*}[htbp]
   \centering
   \includegraphics[width=0.55\textwidth]{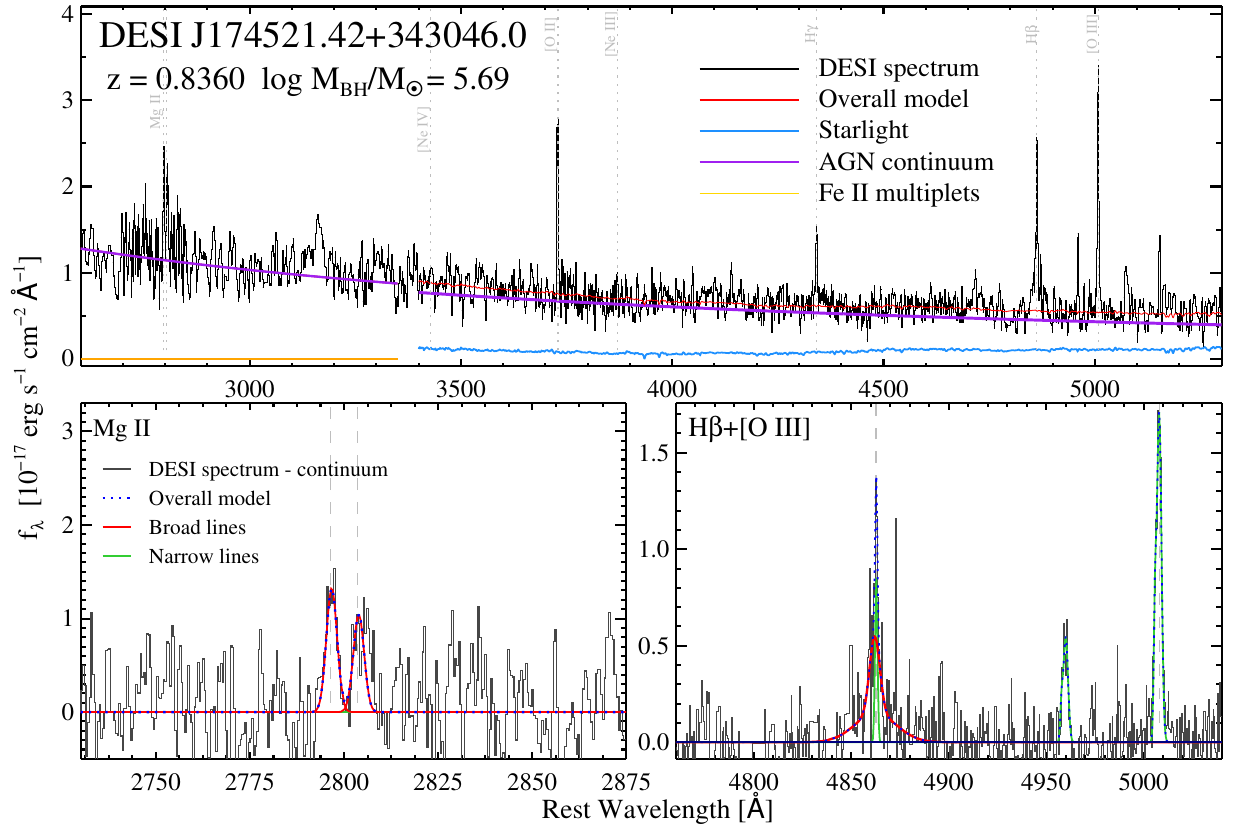}
   \includegraphics[width=0.55\textwidth]{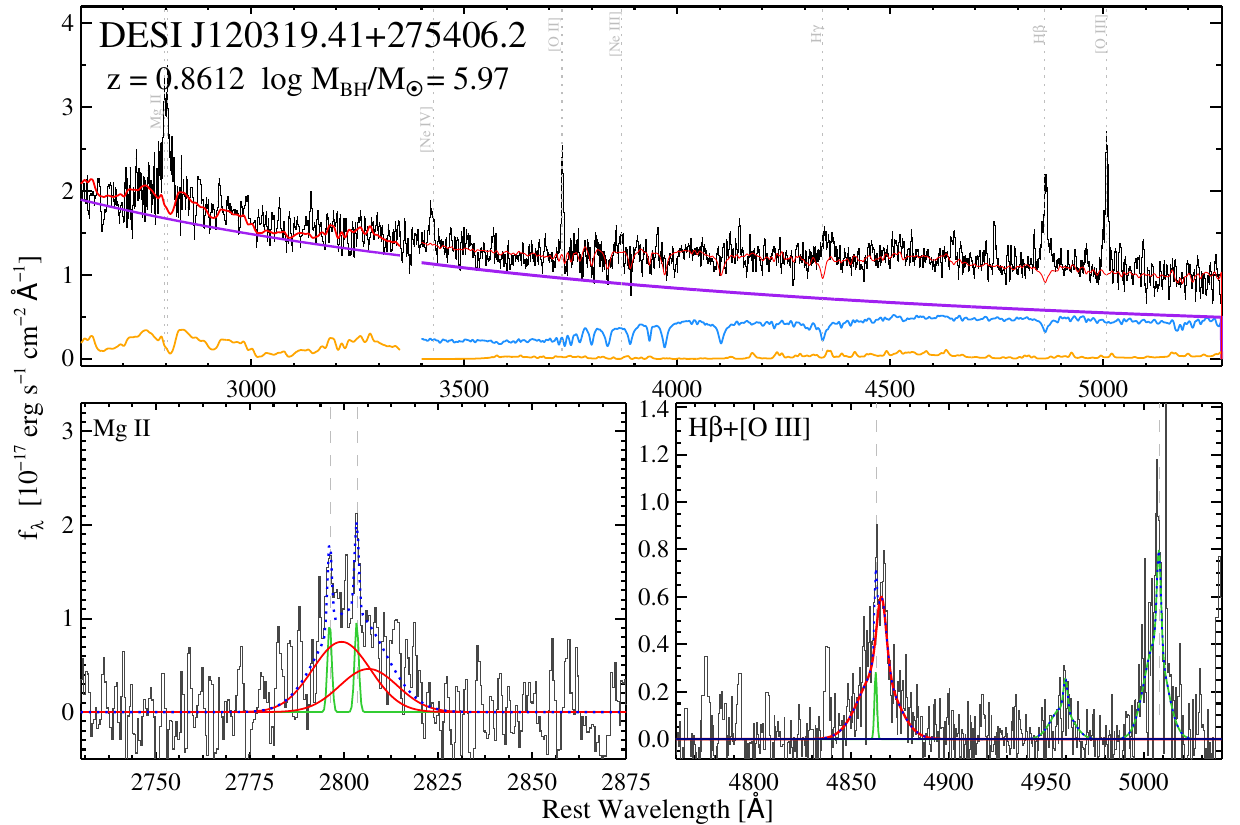}
   \includegraphics[width=0.55\textwidth]{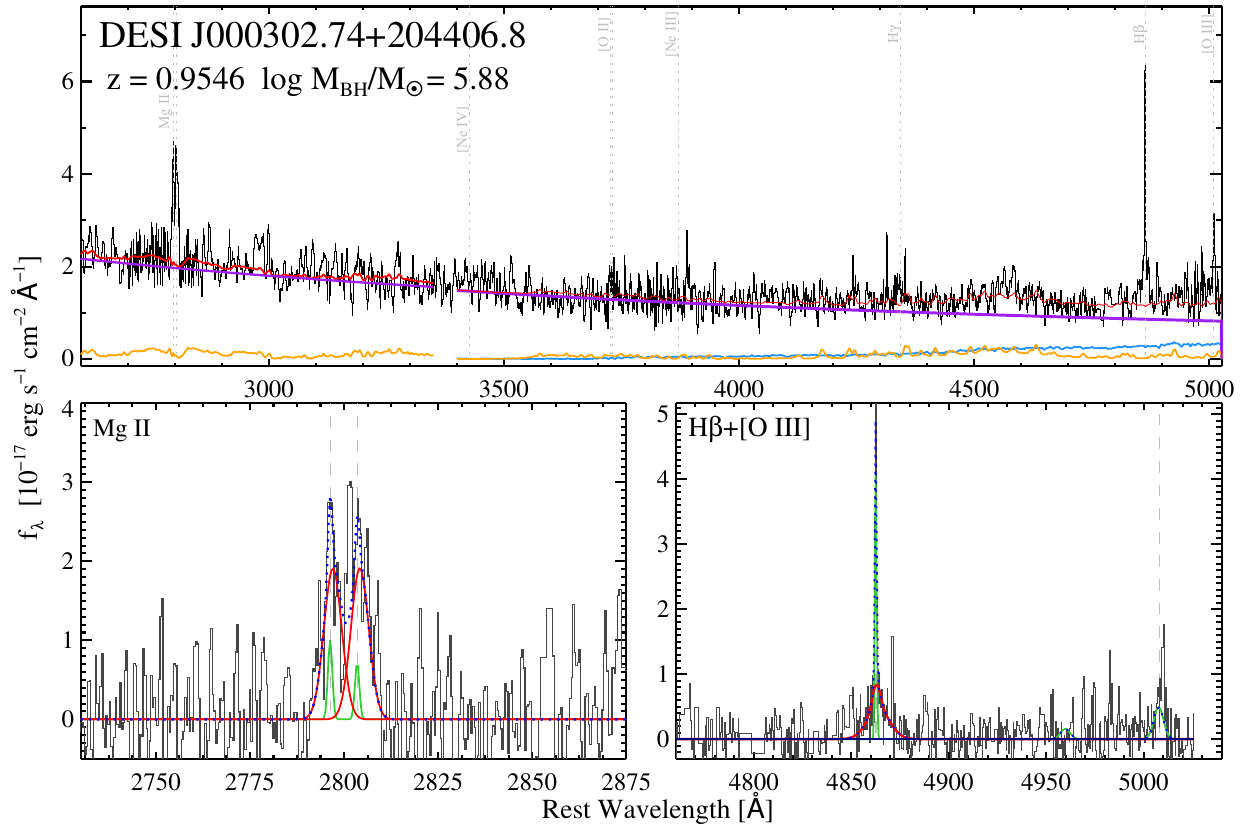}
   \caption{Illustration of the continuum and emission-line fitting for three example IMBH AGNs. 
   For each object, the upper panel shows the continuum decomposition based on the DESI spectrum (black), 
   with the total model overlaid in red. 
   In the optical region, the AGN power-law continuum, host-galaxy starlight, and \feii\ multiplets are shown 
   in purple, blue, and orange, respectively. 
   The UV and optical continuum decompositions are based on independent local fits and are not intended to represent 
   a single continuous global continuum model.
   For the UV local fit, only the empirical local continuum model (purple) and the UV \feii\ component (orange) are displayed;
   the fitted UV continuum is not further decomposed here (see \S~\ref{subsec:mg2fit}).
   Masked bad pixels are excluded from the fits and interpolated only for display.
   For clarity, the DESI spectra are smoothed using a 3-pixel boxcar filter.
   The lower panels display the emission-line fitting for the \mgii\ and \hb+\oiii\ regions.
   The continuum-subtracted spectrum is shown in black,
   the total emission-line fits in blue dotted lines,
   and the broad and narrow components in red and green, respectively.  
   No spectral smoothing was applied in the emission-line panels.
   The figures for the continuum and emission-line decomposition for all sources 
   are available in the journal's online supplementary files.}\label{fig:fig2}
\end{figure*}

\subsubsection{\mgii\ Region \label{subsec:mg2fit}}

Broad \mgii\ emission provides a unique valuable diagnostic for IMBH AGNs, 
particularly at intermediate redshifts where Hydrogen Balmer lines become inaccessible and ambiguous.
Despite its widespread use in luminous quasars, 
systematic studies of \mgii\ emission in IMBH systems have long been lacking,
primarily due to the limited availability of UV-optical with sufficient sensitivity and resolution.
Apart from a few well-studied nearby systems (e.g.,NGC 4395, \citealt{Filippenko1993,Kraemer1999}), 
or a few recently JWST high-$z$ IMBH objects (e.g., GN-z11, \citealt{GNZ11_2024}),
systematic investigations of UV broad-line emission in IMBH AGNs have been largely absent.

In our previous IMBH AGN study \citep{imbh_liu26}, 
we reported the first systematic detection of broad \mgii\ emission in 24 IMBH systems, 
demonstrating that \mgii\ can be robustly identified and modeled in IMBH AGNs.
Building upon this work, 
the enhanced survey depth and spectral resolution of DESI now make large-scale searches 
in IMBH AGN observationally practical over a substantially broader redshift range. 
This enables \mgii\ to serve as both an independent consistency check on 
\hb-based classifications and a key diagnostic for extending IMBH AGNs to higher redshifts.

In this work, we analyze the \mgii\ emission complex and surrounding ultraviolet continuum over 
2700-3300 \AA\ following the general framework established in our previous IMBH AGN studies \citep{imbh_liu26},
and briefly summarize the key steps here.
The fitting strategy is based on that developed for broad-line quasars \citep{mg2wang09},
with adjustments appropriate for lower luminosities and narrower emission lines 
characteristic of IMBH systems.

In the \mgii\ region, the spectral window is relatively narrow and dominated by blended features,
making a full physical decomposition of the ultraviolet continuum difficult to achieve for individual sources.
Our continuum modeling is therefore designed to provide a stable and internally consistent local baseline for the \mgii\ emission-line fitting, 
rather than to isolate all contributing continuum components.
We approximate the underlying continuum with a combination of a single power-law component and UV \feii\ emission.
The power law effectively captures the smooth continuum shape in this wavelength range,
while the UV \feii\ emission is modeled using the semi-empirical template of \citet{uvfe2T06}.
The template is allowed to broaden and shift in velocity to reproduce the observed \feii\ blends.
This simplified treatment is sufficient for robust subtraction of the local continuum beneath the \mgii\ emission.

The \mgii$\lambda\lambda2796,2803$ doublet is modeled by decomposing each line into broad and narrow components.
Within each component, the two lines are assumed to share the same kinematic profile,
with the wavelength separation fixed to the laboratory value and the flux ratio allowed 
to vary between 1:1 and 2:1 \citep{laor1997}.
The broad \mgii\ component is modeled using one or two Gaussians, 
with the addition of a second Gaussian justified by an $F$-test ($P<0.05$).

\begin{deluxetable*}{clcccccccccc}
\centerwidetable
\tablewidth{0pt}
\tabletypesize{\scriptsize}
\tablecaption{The IMBH AGN Sample at $0.5<z\lesssim1$ \label{table:objinfo}}
\tablehead{
\colhead{ID}  & \colhead{Designation} & \colhead{$z_\mathrm{spec}$} & \colhead{DESI Spec} & 
\colhead{NUV} & \colhead{$g$} & \colhead{$r$} & \colhead{$i$} & \colhead{$z$} & 
\colhead{$W1$} & \colhead{$W2$} & \colhead{$E$(B-V)$_\mathrm{MW}$}\\
\colhead{(1)} & \colhead{(2)} & \colhead{(3)} & \colhead{(4)} & 
\colhead{(5)} & \colhead{(6)} & \colhead{(7)} & \colhead{(8)} & \colhead{(9)} &
\colhead{(10)} & \colhead{(11)} & \colhead{(12)}
}
\startdata
   1 & J000103.99$+$144838.0 & 0.7137 & 39628137566306528-main-dark & 23.57 & 22.63 & 22.22 & 21.70 & 21.49 & 19.00 & 18.51 & 0.044\\
   2 & J000302.74$+$204406.8 & 0.9546 & 39628275873484792-main-dark &       & 21.51 & 21.25 & 20.93 & 20.69 & 19.13 & 18.88 & 0.054\\
   3 & J000903.27$-$075942.9 & 0.5548 & 39627592000607372-main-dark &       & 21.81 & 21.23 & 20.84 & 20.60 & 19.80 & 19.90 & 0.034\\
   4 & J001234.88$+$101906.0 & 0.5770 & 39628031328782313-main-dark &       & 22.14 & 20.92 & 20.16 & 20.00 & 17.97 & 17.57 & 0.091\\
   5 & J005908.94$+$004908.4 & 0.5903 & 39627803095729383-main-dark & 22.76 & 21.58 & 20.27 & 19.55 & 19.15 & 16.77 & 16.34 & 0.025\\
   6 & J010816.42$+$203127.0 & 0.6610 & 39628270471217208-main-dark &       & 21.70 & 21.21 & 20.83 & 20.68 & 19.04 & 19.12 & 0.045\\
   7 & J012731.59$+$200539.9 & 0.8660 & 39628259180153054-main-dark &       & 22.02 & 21.70 & 21.34 & 21.00 & 19.24 & 19.15 & 0.044\\
   8 & J012820.94$+$202514.8 & 0.5335 & 39628270546719373-main-dark & 21.66 & 21.54 & 21.13 & 21.06 & 20.76 & 19.78 & 19.46 & 0.044\\
   9 & J014033.07$+$183846.0 & 0.5574 & 39628230671468056-main-dark & 22.28 & 21.12 & 20.41 & 20.13 & 19.81 & 18.14 & 18.02 & 0.053\\
  10 & J020003.38$-$000223.4 & 0.5437 & 39627785232187929-main-dark &       & 20.95 & 19.90 & 19.33 & 18.98 & 17.56 & 17.39 & 0.028\\
\enddata
\vspace{0.5em} 
\textbf{Notes:} \\
Col. (1): Identification number assigned in this paper.
Col. (2): Official DESI designation in J2000.
Col. (3): Redshift measured by the DESI pipeline.
Col. (4): The DESI spectrum used for fitting (Target-Survey-Program). 
Col. (5): GALEX NUV magnitude, using the MIS measurement when available and the AIS measurement otherwise.
Cols. (6)-(9): DESI $g$, $r$, $i$, $z$ magnitudes. 
Cols. (10)-(11): WISE W1 and W2 magnitudes. 
Col. (12): Galactic color excess, $E$(B−V)$_\mathrm{MW}$, from the DESI photometric catalog. 
The listed photometric magnitudes are not corrected for Galactic extinction. 
(This table is available in its entirety in a machine-readable form in the online journal.
A portion is shown here for guidance regarding its form and content.)
\end{deluxetable*}

\subsection{Robust Identification of Broad Lines in Low-mass Regime \label{subsec:creteria}}

Compared to broad \ha, broad \hb\ emission is intrinsically weaker and more susceptible 
to internal extinction from dust within the host galaxy.
These factors make the broad \hb\ component fainter and 
thus more challenging to detect and decompose,
given in the low per-pixel S/N of the spectra (average $\sim 4$ ).
Rather than prioritizing maximum completeness, 
we adopt a conservative selection strategy aimed at identifying a reliable sample of IMBH AGN candidates,
for which the presence of broad-line emission is supported by quantitative fitting,
visual inspection, and consistency across multiple emission lines.

In principle, broad \mgii\ emission could serve as an alternative diagnostic at these redshifts.
However, we do not use \mgii\ as the primary selection criterion.
First, BH masses are ultimately anchored to the \hb\ $R-L$ relation,
while \mgii\ still lacks a well-established reverberation-mapping calibration in this regime.
Moreover, at the low-mass (or low-FWHM) end relevant for IMBHs,
the correlation between \mgii\ and \hb\ line widths remains poorly constrained 
and may deviate systematically (see \citealt{imbh_liu26}, Figure~7).
introducing additional uncertainty in mass estimates based on \mgii.
We therefore employ \mgii\ only as a supporting diagnostic in this work.
A dedicated investigation of the \mgii-\hb\ relationship using higher-S/N sources 
will be presented in a forthcoming work.


Our identification of broad \hb\ emission is based on a joint evaluation of spectral decomposition results, 
statistical significance, and physical consistency across multiple emission lines.
The emission-line decomposition described in \S~2.2.1 explores multiple plausible descriptions of the observed \hb\ profile, 
including different treatments of the narrow-line component motivated by the diversity of observed narrow-line kinematics. 
These candidate models are then evaluated using both quantitative criteria and physical consistency checks.

We require a statistically significant broad component from spectral fitting 
using quantitative criteria adapted from those employed in our previous \ha-based IMBH AGN selections 
\citep{imbh_dong12, imbh_liu18, imbh_liu26}, 
but modified to account for the intrinsically weaker broad \hb\ emission.
After the emission-line fitting converges to an acceptable solution ($\chisq < 1.5$),
a broad \hb\ component is considered statistically significant if it satisfies the following criteria:
$P_{F-\mathrm{test}} < 0.05$, 
S/N$_{\hb^\mathrm{b}} \geqslant 3$, 
$h_\mathrm{B} \geqslant 2$ rms,
and FWHM$_{\hb^\mathrm{b}} >$ FWHM$_\mathrm{\oiii}$.
Here $h_\mathrm{B}$ is defined as the peak height of the broad \hb\ component above the local continuum residuals, 
measured in units of the rms of the continuum-subtracted spectrum.

When applying the above criterion, 
we find that the relationship between the observed \hb\ and \oiii\ profiles varies significantly across the sample.
In some sources, the overall \oiii\ profile in some sources exhibits substantial asymmetry, 
broad wings, or even globally blueshifted kinematics relative to the systemic velocity,
resulting in line profiles that differ significantly from the observed narrow \hb\ component 
(see representative examples in Figure~\ref{fig:fig1}).
We therefore adopt the \oiii\ core component as the operational reference for the narrow-line width in the above criterion, 
while further evaluating the physical plausibility of the inferred broad \hb\ component through consistency checks involving other emission lines and visual inspection. 
In particular, we examine the morphology and kinematics of other emission lines, 
including \mgii, \oiii, \oii, and the optical \feii\ multiplets. 
The presence of additional BLR-related features, 
such as broad \mgii\ emission and/or prominent UV or optical \feii\ complexes, 
provides supporting evidence for the broad \hb\ identification.

This combined approach allows us to balance robustness and completeness 
in the identification of broad-line IMBH AGN candidates in the low-S/N regime probed by this work. 
The selected broad-line AGN candidates are then used to estimate BH masses.

\begin{deluxetable*}{cCCCCCCCCC}
\centerwidetable
\tabletypesize{\scriptsize}
\tablecaption{Emission-line Measurements \label{table:lineinfo}}
\tablehead{
\colhead{ID} & \multicolumn{6}{c}{log Flux} & \multicolumn{3}{c}{FWHM} \\
\colhead{} & \multicolumn{6}{c}{(erg~s$^{-1}$~cm$^{-2}$)} & \multicolumn{3}{c}{({\kms})} \\
\cline{2-7} \\[-11pt] \cline{8-10} \\[-9pt] \cline{8-10} 
\colhead{}  & 
\colhead{[O\,{\scriptsize II}] $\lambda3727$} & \colhead{Fe\,{\scriptsize II}\,$\lambda4570$} & 
\colhead{H$\beta^\mathrm{n}$} &  \colhead{[O\,{\scriptsize III}] $\lambda5007$} &
\colhead{Mg\,{\scriptsize II}$^\mathrm{b}$} & \colhead{H$\beta^\mathrm{b}$} &
\colhead{Mg\,{\scriptsize II}$^\mathrm{b}$} & \colhead{H$\beta^\mathrm{b}$} & \colhead{[O\,{\scriptsize III}]} \\
\colhead{(1)} & 
\colhead{(2)} & \colhead{(3)} & \colhead{(4)} & \colhead{(5)} & \colhead{(6)} & \colhead{(7)} & 
\colhead{(8)} & \colhead{(9)} & \colhead{(10)} 
}
\startdata
  1 & -16.44$\pm$0.07 & -16.37$\pm$0.01 & -15.98$\pm$0.05 & -15.77$\pm$0.02 & -15.67$\pm$0.01 & -15.72$\pm$0.05 & 593$\pm$34 & 527$\pm$159 & 428$\pm$68\\
  2 & <-16.32 & -15.43$\pm$0.01 & -16.01$\pm$0.09 & -16.22$\pm$0.06 & -15.64$\pm$0.01 & -15.77$\pm$0.05 & 593$\pm$38 & 494$\pm$223 & 346$\pm$31\\
  3 & -16.14$\pm$0.07 & -16.23$\pm$0.01 & <-16.98 & -15.81$\pm$0.02 &  & -15.89$\pm$0.02 & 465$\pm$0 & 845$\pm$12 & 230$\pm$19\\
  4 & -16.32$\pm$0.10 & -15.62$\pm$0.01 & <-16.40 & -15.79$\pm$0.04 &  & -15.62$\pm$0.03 & 465$\pm$0 & 702$\pm$14 & 295$\pm$45\\
  5 & -15.86$\pm$0.06 & -15.79$\pm$0.01 &  & -15.74$\pm$0.02 &  & -15.58$\pm$0.01 & 465$\pm$0 & 690$\pm$29 & 450$\pm$59\\
  6 & -15.89$\pm$0.10 & -16.54$\pm$0.01 & -16.12$\pm$0.04 & -16.33$\pm$0.05 & -15.60$\pm$0.04 & -15.88$\pm$0.12 & 831$\pm$244 & 697$\pm$246 & 402$\pm$78\\
  7 & <-16.25 & -15.61$\pm$0.01 & -17.15$\pm$0.13 & -16.51$\pm$0.11 & -15.66$\pm$0.23 & -16.04$\pm$0.01 & 1495$\pm$700 & 601$\pm$41 & 313$\pm$83\\
  8 & <-16.20 & -15.71$\pm$0.01 & <-16.52 & -16.00$\pm$0.02 & -15.85$\pm$0.17 & -15.69$\pm$0.05 & 662$\pm$665 & 740$\pm$189 & 191$\pm$45\\
  9 & <-15.53 & -16.06$\pm$0.01 & -16.41$\pm$0.05 & -15.71$\pm$0.02 & -15.44$\pm$0.04 & -15.64$\pm$0.10 & 826$\pm$126 & 610$\pm$202 & 851$\pm$262\\
 10 & -16.03$\pm$0.07 & <-16.80 & -16.32$\pm$0.12 & -15.81$\pm$0.03 & -15.22$\pm$0.02 & -15.79$\pm$0.05 & 1415$\pm$129 & 801$\pm$113 & 416$\pm$38\\
\enddata
\vspace{0.5em} 
\textbf{Notes:} \\
Col. (1): Identification number assigned in this paper.
Cols. (2)-(7): Emission-line fluxes. 
Note that the fluxes are observed values with no NLR or BLR extinction correction applied.
The superscripts ``n'' and ``b'' refer to the narrow and broad components of the line, respectively.
Mg\,{\scriptsize II}$^{b}$ flux in Col.(6) corresponds to the combined flux of the doublet \mgii$\lambda\lambda2796,2803$. 
The measured emission-line fluxes are regarded to be reliable detections if they have significance greater than 3$\sigma$,
or else $3\sigma$ values are adopted as the upper limit. 
Cols. (8)-(10): Line widths, corrected for instrumental broadening.
(This table is available in its entirety in a machine-readable form in the online journal.
A portion is shown here for guidance regarding its form and content.)
\end{deluxetable*}


\begin{deluxetable}{CCCCCCCC}
 \centerwidetable
 \movetableright=0in
\tablewidth{0pt}
\tabletypesize{\footnotesize}
\tablecaption{Physical Properties of the Sample\label{table:physinfo}}
\tablehead{
\colhead{ID} & 
\colhead{log\,$L_{\hb^\mathrm{b}}$} & \colhead{log\,$\lambda L_{3000}$} & \colhead{log\,$\lambda L5100$} & \colhead{log\,$L_\mathrm{bol}$} &
\colhead{$R_\mathrm{Fe II}$} & \colhead{log\,\mbh } & \colhead{log\,\redd} \\ 
\colhead{}   & 
\colhead{(erg~s$^{-1}$)} & \colhead{(erg~s$^{-1}$)} & \colhead{(erg~s$^{-1}$)} & \colhead{(erg~s$^{-1}$)} & 
\colhead{} & \colhead{(\msun)} & \colhead{} \\
\colhead{(1)} & \colhead{(2)} & \colhead{(3)} & \colhead{(4)} & \colhead{(5)} &
\colhead{(6)} & \colhead{(7)} & \colhead{(8)}  
}
\startdata
  1 & 41.64$\pm$ 0.05 & 43.70$\pm$ 0.01 & 43.69$\pm$ 0.01 & 44.70$\pm$ 0.03 &  0.21$\pm$0.02 & 5.80$\pm$0.26 &  0.81$\pm$0.27\\
  2 & 41.90$\pm$ 0.05 & 44.40$\pm$ 0.01 & 44.28$\pm$ 0.01 & 44.94$\pm$ 0.04 &  2.01$\pm$0.21 & 5.88$\pm$0.39 &  0.95$\pm$0.40\\
  3 & 41.20$\pm$ 0.02 & 43.64$\pm$ 0.01 & 43.44$\pm$ 0.01 & 44.16$\pm$ 0.02 &  0.41$\pm$0.03 & 5.96$\pm$0.02 &  0.10$\pm$0.03\\
  4 & 41.51$\pm$ 0.03 & 43.64$\pm$ 0.02 & 43.94$\pm$ 0.01 & 44.45$\pm$ 0.03 &  1.02$\pm$0.05 & 5.98$\pm$0.02 &  0.38$\pm$0.04\\
  5 & 41.58$\pm$ 0.01 & 43.69$\pm$ 0.01 & 44.10$\pm$ 0.01 & 44.48$\pm$ 0.01 &  0.65$\pm$0.02 & 6.00$\pm$0.04 &  0.39$\pm$0.04\\
  6 & 41.40$\pm$ 0.12 & 43.91$\pm$ 0.01 & 43.92$\pm$ 0.01 & 44.50$\pm$ 0.07 &  0.20$\pm$0.02 & 5.90$\pm$0.31 &  0.49$\pm$0.32\\
  7 & 41.53$\pm$ 0.01 & 44.15$\pm$ 0.01 & 43.98$\pm$ 0.01 & 44.47$\pm$ 0.01 &  2.41$\pm$0.15 & 5.85$\pm$0.06 &  0.52$\pm$0.06\\
  8 & 41.36$\pm$ 0.05 & 43.68$\pm$ 0.01 & 43.68$\pm$ 0.01 & 44.32$\pm$ 0.04 &  0.91$\pm$0.05 & 5.94$\pm$0.22 &  0.28$\pm$0.23\\
  9 & 41.46$\pm$ 0.10 & 44.01$\pm$ 0.01 & 43.96$\pm$ 0.01 & 44.43$\pm$ 0.08 &  0.35$\pm$0.01 & 5.82$\pm$0.29 &  0.51$\pm$0.30\\
 10 & 41.28$\pm$ 0.05 & 43.61$\pm$ 0.02 & 43.71$\pm$ 0.01 & 44.32$\pm$ 0.04 &                & 5.96$\pm$0.13 &  0.26$\pm$0.13\\
\enddata
\vspace{0.5em}
{\textbf {\footnotesize Notes:}}
{
Column (1): Identification number assigned in this paper.
Column (2): Luminosity of broad \hb.
Column (3): Monochromatic luminosity at 3000\,\AA.
Column (4): Monochromatic luminosity at 5100\,\AA, derived from the decomposed power-law continuum.
Column (5): Bolometric luminosity.
Column (6): Optical \feii\ strength, defined as $R_{\feii} \equiv EW_{\feii}/EW_{\hb}$, where EW$_{\feii}$ is measured over 4,434--4,4684 \AA.
Column (7): Virial BH mass estimate using the broad \hb-based calibration from \citet{gh05}.
The quoted uncertainties include measurement uncertainties propagated from the emission-line fitting, 
including systematic components associated with spectral decomposition. 
They do not include the intrinsic scatter of the adopted virial scaling relations (typically $\sim$ 0.4--0.5 dex). 
Column (8): Eddington ratio.
}
\end{deluxetable}

\section{The Sample of Broad-line IMBHs at $0.5<z<1$}

Having identified a robust set of broad-line AGN candidates 
based on \hb\ and auxiliary diagnostics, 
we proceed to estimate BH masses and construct 
the final sample of IMBHs at $0.5<z<1$. 
In this section, we first describe the virial mass 
estimation method adopted in this work, 
followed by the definition of the final IMBH sample 
and an overview of its basic properties.

\subsection{BH Mass \label{subsec:BHmass} and Eddington Ratio}

Broad \hb\ emission is widely regarded as the most reliable tracer 
of virialized gas in the broad-line region,
providing the most robust single-epoch BH mass estimates 
among commonly used optical lines.
Also, it is supported by the largest number of reverberation-mapping studies,
which have established well-calibrated radius-luminosity ($R-L$) 
relations across a broad range of luminosities and BH masses.
Accordingly, we adopt an \hb-based virial calibration to estimate BH masses for our sample.

Specifically, we use the \citet{gh05} \hb\ calibration, 
which relates the BH mass to the FWHM and luminosities of broad \hb.
This calibration, together with its \ha-based counterpart, 
has been widely employed in previous studies of IMBH AGN studies.
Adopting this formalism ensures consistency with earlier IMBH samples and 
facilitates direct comparison across different works in the literature.
We note that more recent calibrations, such as those proposed by \citet{hk15},
incorporate updated $R-L$ relations and explicitly account for host-galaxy bulge properties,
yielding BH masses that are systematically larger than those 
from \citet{gh05} by an average of $\sim$ 0.33 dex.
For the purposes of this work, 
we retain the \citet{gh05} calibration as our primary mass scale.

Estimating the bolometric luminosity is essential for characterizing the accretion properties of IMBH AGNs.
In optical AGN studies, \lbol\ is commonly estimated from the monochromatic continuum luminosity at 5100,\AA. 
For IMBH AGNs, however, reliable measurements of the nuclear optical continuum can be challenging because 
the observed spectra may contain substantial host-galaxy starlight.
Even luminous and blue AGNs can have a non-negligible host-galaxy contribution \citep{vandenberk01}.
This issue is particularly relevant for our sample because the fixed 1.5\arcsec\ DESI fiber corresponds 
to a relatively large physical scale at $0.5<z<1$, 
allowing substantial host-galaxy light to enter the aperture. 
Moreover, the relatively modest S/N of our spectra makes the stellar absorption features difficult to constrain in many objects, 
introducing systematic uncertainties into the separation of the nuclear power-law continuum and host-galaxy starlight.

We therefore infer \lfive\ from the broad \hb\ luminosity using the empirical relation of \citet{gh05}, 
and adopt this \hb-based estimate as our fiducial continuum luminosity. 
We then calculate the bolometric luminosity using a fixed bolometric correction of $\lbol=9.8\,\lfive$ \citep{Mclure04}.
We note that the bolometric correction may depend on luminosity, black hole mass, and accretion rate \citep[e.g.,][]{netzer19}; 
however, existing empirical studies also suggest that the optical bolometric correction can remain approximately constant over 
a broad luminosity range \citep[e.g.,][]{duras20}. 
Since no empirical calibration is currently available specifically for the low-mass, high-Eddington-ratio regime probed here, 
we do not adopt a luminosity-dependent correction that would require extrapolation beyond its calibrated parameter space. 
The Eddington ratio is then calculated as \redd, where $\ledd = 1.26 \times 10^{38}\,(\mbh/\msun)$ \ergs.

As a consistency check, we also derive \lfive\ directly from the decomposed AGN power-law component. 
The power-law component contributes substantially to the fitted continuum: 
its fraction at 4200,\AA\ has a median of 0.68, respectively, with a minimum of 0.46, and exceeds 60\% in 74.5\% of the sources. 
However, a high fitted power-law fraction does not necessarily imply that the nuclear continuum is accurately isolated. 
When the S/N is limited, weak or unresolved stellar absorption features can make the stellar component difficult to constrain, 
potentially causing some host-galaxy light to be absorbed into the fitted power-law component.
The directly measured \lfive\ values are on average $\sim0.39$ dex higher than those inferred from broad \hb\ in the present $0.5<z<1$ sample. 
Applying the same comparison to the $z<0.6$ IMBH sample yields a substantially smaller mean difference of $\sim0.15$ dex.
The larger discrepancy at higher redshift may reflect, at least in part, 
the increased difficulty of separating host-galaxy and nuclear continuum components in the present spectra. 




\subsection{The Final IMBH Sample \label{subsec:finalsample}}

Applying the BH mass estimates derived in \S~\ref{subsec:BHmass}, 
we define our final IMBH AGN sample by adopting 
an upper mass threshold of \mbh$\leqslant10^{6}$\msun,
which is commonly used to delineate the IMBH 
regime in the literature (e.g., \citealt{Greene2020}).
Under this criterion, 
we identify 98 broad-line IMBH AGN candidates within the redshift range $0.5<z<1$.

Table~\ref{table:objinfo} summarizes the basic source information, 
including source designation, coordinates, spectroscopic redshift, 
DESI spectrum identifier, and multiwavelength photometry.
The measured spectral properties are presented in Table~\ref{table:lineinfo}, 
while the derived physical properties are summarized in Table~\ref{table:physinfo}.
Among the sample, 47 objects exhibit identifiable broad \mgii\ emission. 

Figure~\ref{fig:fig2} presents three representative examples of the spectral fitting 
for sources in the final IMBH AGN sample.
We select two relatively high-redshift objects with good detection of broad \mgii\ emission.
The third source, J0030$+$2044, is the highest-redshift object in our sample.
Although the \hb\ region in this spectrum has relatively low S/N, 
the presence of prominent optical \feii\ emission complexes and robust \mgii\ lines provides 
independent and compelling support for the identification of a genuine broad \hb\ component.

Our IMBH AGN sample is selected based on the detection of broad \hb\ emission,
for which the line flux and FWHM constitute the two most fundamental observational quantities.
Notably, no explicit hard cuts on either the broad-line flux or line width were 
imposed during the sample selection.
As a result, the final sample itself provides an empirical basis for examining the range of 
broad \hb\ fluxes and FWHM that can be reliably identified in DESI spectra.
Such an assessment is essential for informing future evaluations of the effective sensitivity and completeness of broad-line detection,
and also offers useful guidance for further searches for IMBH AGNs at other redshift intervals.

\begin{figure}[htbp]
   \centering
   \includegraphics[width=0.495\textwidth]{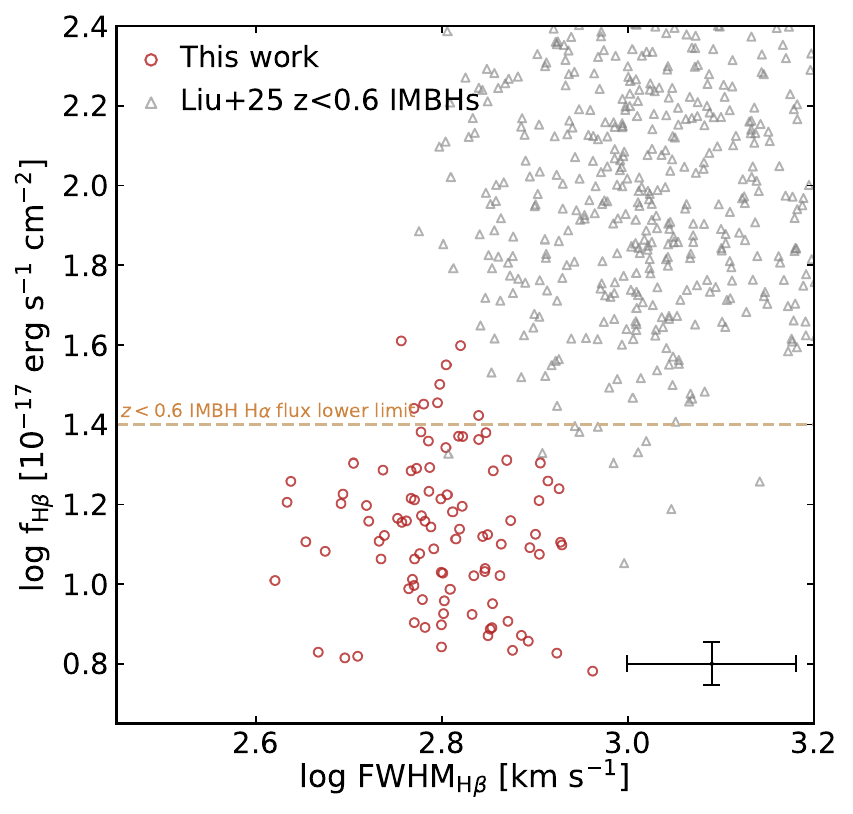}
   \caption{Broad \hb\ flux versus FWHM for the final IMBH AGN sample 
   at $0.5<z\lesssim1$ (black circles). 
   For comparison, a subset of 585 IMBH AGNs at $z\lesssim0.6$ drawn from 
   \citet{imbh_liu26} is shown (light blue triangles), 
   selected to have reliable broad \hb\ measurements and BH masses of \mbh$\lesssim10^{6}$\msun,
   using the same \hb-based mass calibration as adopted in this work.
   The orange dashed line indicates the effective broad \ha\ flux boundary 
   of the SDSS-based IMBH AGN selection. 
   A representative 1-$\sigma$ error bar is shown (lower-right), 
   representing the total uncertainty ($\sigma_\mathrm{total}$; see \citealt{imbh_dong12}), 
   which combines contributions from statistical noise ($\sigma_\mathrm{n}$) 
   and subtraction of nearby narrow lines ($\sigma_\mathrm{NL_{sub}}$).
   For this sample, $\sigma_\mathrm{total} \approx 2.7 \sigma_\mathrm{n}$ for flux and 
   $\approx 3.5 \sigma_\mathrm{n}$ for FWHM. 
   This figure illustrates the improved sensitivity of DESI spectroscopy
   for detecting weak and relatively narrow broad-line emission in IMBH AGNs.}\label{fig:fig3}
\end{figure}

Figure~\ref{fig:fig3} presents the distribution of broad \hb\ flux as 
a function of FWHM for the final IMBH sample at $0.5<z<1$.
The detected sources span a broad \hb\ flux range of $6-40\times10^{-17}$ erg s$^{-1}$ cm$^{-2}$
and FWHMs from $\sim 417-914$\kms. 
To facilitate a direct comparison with $z<0.6$ IMBH AGNs \citep{imbh_liu26}, 
we additionally include a subsample of 558 objects selected from the $z<0.6$ IMBH AGN sample.
Since that sample was primarily selected based on broad \ha\ emission,
we re-examine the \hb\ line fits for all 930 sources 
and apply the same \hb-based mass estimator used in this work. 
We then select objects with reliable broad \hb\ measurements and derived BH mass with log\,$\mbh/\msun<6$.
This procedure yields a comparison sample of 585 IMBH AGNs at $z<0.6$,
enabling a consistent and homogeneous comparison across redshift.

Compared with the SDSS-based low-redshift sample, 
the DESI-selected IMBH AGNs extend to systematically lower broad \hb\ fluxes by approximately 0.5 dex, 
while also reaching narrower measured broad-\hb\ FWHMs.
The lower flux limit primarily reflects the greater observational depth of DESI, 
which enhances the detectability of weak broad-line components. 
The apparent extension toward smaller FWHMs, however, should be interpreted with some caution.
While the higher spectral resolution of DESI helps resolve relatively narrow broad-line components, 
the higher-redshift DESI spectra generally have lower effective S/N around the observed \hb\ region,
making the broad-line wings more difficult to constrain and potentially biasing some FWHM measurements toward smaller values.
Moreover, the low-redshift comparison sample was constructed with relatively stringent broad-line criteria, 
including requirements on the broad-line width relative to the narrow-line width, 
which may preferentially exclude the narrowest broad-line candidates.
Therefore, the difference in the minimum measured FWHM between the two samples should not be interpreted as direct evidence 
for an intrinsic difference in broad-line kinematics.

For reference, we also indicate in Figure~\ref{fig:fig3} the effective broad 
\ha\ flux boundary of the SDSS-based IMBH AGN selection using a brown dashed line,
which lies at a comparable level to the lower boundary of the SDSS-based \hb\ measurements.
This indicates that DESI achieves a $\sim0.5$ dex gain in sensitivity for broad-line detection relative to SDSS.
Importantly, in the observed flux--FWHM plane, the limiting broad-line flux is set primarily 
by the spectral depth and line detectability, rather than by redshift itself.
At a given observed flux limit, however, sources at different redshifts
correspond to different intrinsic luminosity limits and therefore probe
different regions of BH mass and Eddington-ratio parameter space.
Thus, if DESI-quality spectra were available for low-redshift systems, they
would likely enable the identification of IMBH AGNs down to lower BH
masses and/or lower Eddington ratios than previously accessible.

\begin{figure*}[htbp]
   \centering
   \includegraphics[width=0.8\textwidth]{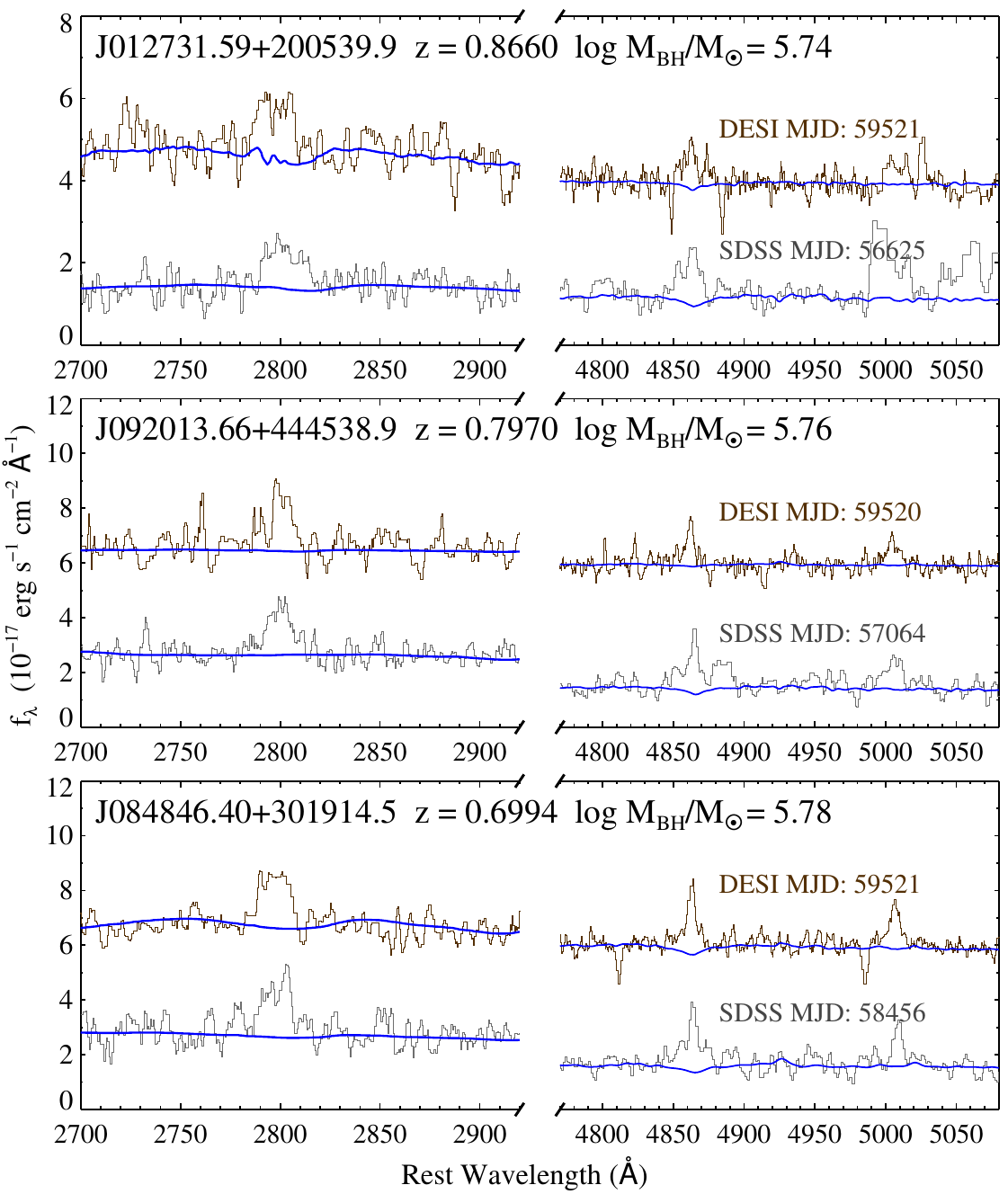}
   \caption{Comparison between DESI and SDSS spectra for three representative IMBH AGNs observed by both surveys. 
   For each object, the left panel shows the \mgii\ region, and the right panel presents the \hb+\oiii\ region. 
   The DESI and SDSS spectra are shown together for direct comparison,
   with the DESI spectra vertically offset for clarity.
   The best-fitting continuum models are overlaid as blue solid line for the corresponding wavelength ranges. }\label{fig:fig4}
\end{figure*}

A subset of our IMBH AGN sample has also been previously observed by SDSS,
providing an independent check on the robustness of our broad-line identification.
Among the 98 IMBH AGN candidates, 12 objects have available SDSS spectra.
Broad-line emission is detected in 8 of these sources in the SDSS data:
6 objects show detectable broad \hb\ emission, 
while the remaining 2 exhibit broad \mgii\ emission but lack clear broad \hb\ detection.
The remaining four objects do not show detectable broad-line features in the SDSS spectra.
We inspected these cases individually to assess whether the non-detections are related to data quality 
or intrinsic variability.
None of these four sources shows a clear broad \mgii\ detection in the DESI spectra either.
The broad \hb\ emission detected in the DESI spectra is relatively weak and, for these sources, 
is already close to the detection threshold.
The corresponding SDSS spectra generally have even lower S/N in the relevant wavelength regions, 
making these broad-line components difficult to distinguish.
The continuum levels do not show obvious changes between the SDSS and DESI epochs. 
One source, J1240+3721, does show an apparent strengthening of the broad \hb emission in the DESI spectrum relative to SDSS, 
although the relatively low S/N prevents us from reliably quantifying the amplitude of this variation. 
Overall, the SDSS non-detections appear to be primarily related to the lower data quality, 
although broad-line variability may contribute in individual cases.

Figure~\ref{fig:fig4} presents a direct comparison between the DESI and SDSS spectra for three representative objects.
For each object, we show the \mgii\ region and the \hb+\oiii\ region, 
enabling a direct visual comparison fo the broad-line features across different epochs and instruments.
Overlaid on each spectrum are the best-fitting continuum models appropriate for the corresponding wavelength ranges.
In both the DESI and SDSS spectra, both broad \mgii\ and \hb\ emission lines are detected for all three sources.

\section{Results}

\subsection{Sample Distributions \label{subsec:distributions}}

\begin{figure}[htbp]
   \centering
   \includegraphics[width=0.495\textwidth]{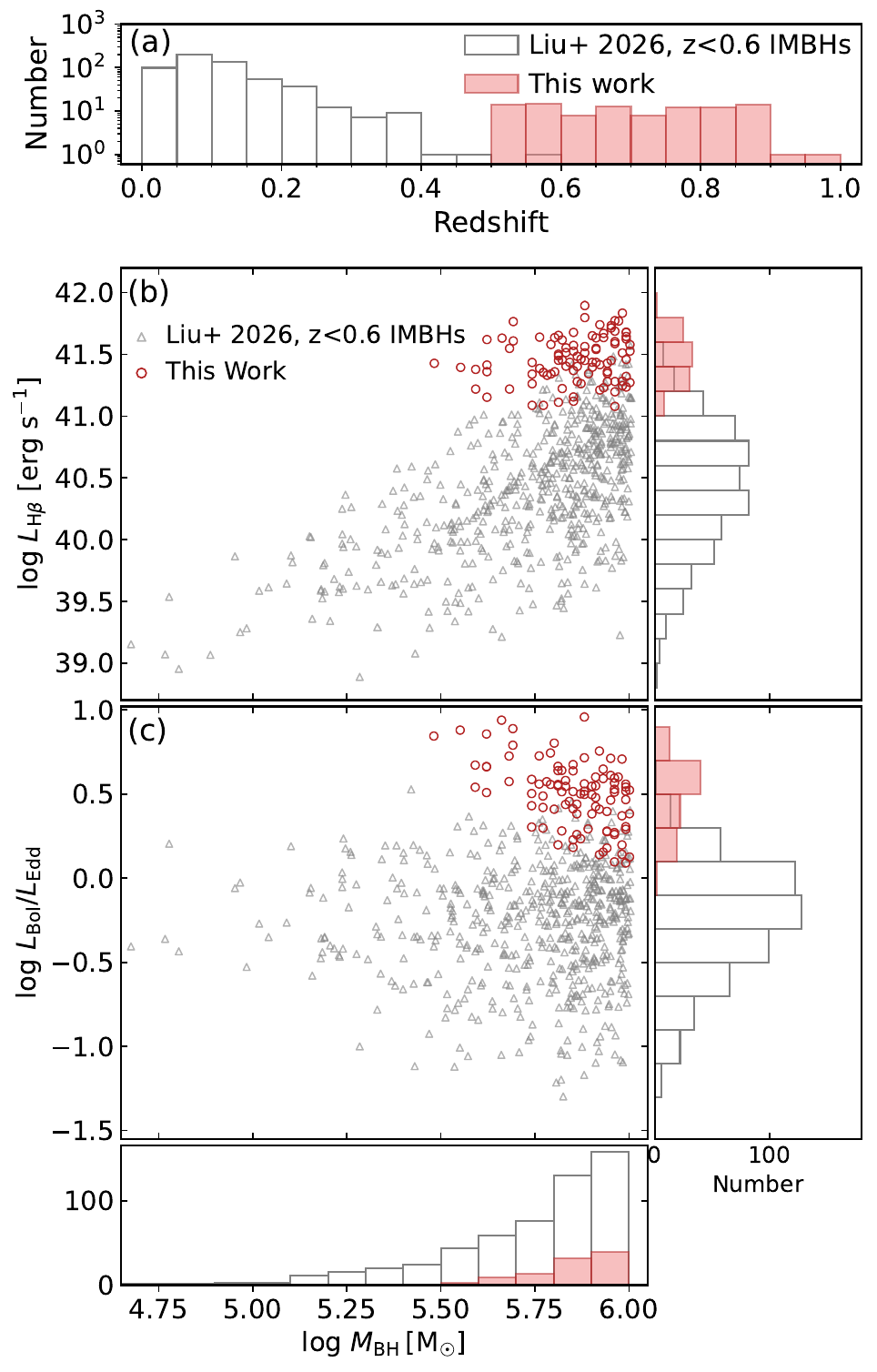}
   \caption{Distributions of key physical properties for the 98 broad-\hb\ IMBH AGNs at $0.5<z\lesssim1$ identified 
   in this work (red), overlaid with the $z<0.6$ IMBH AGNs with robust broad-\hb\ detections from \citet{imbh_liu26} (gray). 
   (a) Redshift distribution. The $0.5<z<1$ sample is shown by the red histogram, 
   and the $z<0.6$ comparison sample is shown by the gray open histogram.
   (b) Distribution in the log\mbh-log$L_\mathrm{\hb}$ plane.
   (c) Distribution in the log\mbh-log \redd\ plane.
   In panel (b) and (c), red open circles and gray open triangles denote the
   $0.5<z<1$ sample and the $z<0.6$ comparison sample, respectively.
   The horizontal marginal histogram shows the \mbh\ distributions, 
   while the vertical marginal histograms show the broad-\hb\ luminosity and Eddington-ratio distributions, respectively.}\label{fig:fig5_sampledistri}
\end{figure}

Having established the final IMBH AGN sample, 
we now examine the global distributions of physical properties.
Figure~\ref{fig:fig5_sampledistri} presents the redshift distribution and the locations of the 98 broad-line IMBH AGNs in the log\,\mbh-log\hb\ and log\mbh-log\redd\ planes.
For comparison, we also show the corresponding distributions for the 585 $z<0.6$ IMBH AGNs with robust broad \hb\ detections from \citet{imbh_liu26} (see \S\,\ref{subsec:finalsample} for details).

The redshift distribution spans $0.5<z<1$,
as required for the selection of broad \hb\ IMBH AGNs from DESI spectra.
Compared with the $z<0.6$ sample, 
the present work extends the systematic identification of broad-line IMBH AGNs to substantially higher redshifts, 
providing a sample in the intermediate-redshift regime.

The BH mass distribution of the present sample are bounded above by the selection 
criterion at $\mbh=10^6$~\msun\ and extend down to $\mbh=10^{5.5}$\msun.
As shown in Figure~\ref{fig:fig5_sampledistri}, 
$0.5<z<1$ sample is more concentrated toward the upper end of the IMBH regime than the $z<0.6$ comparison sample.
This difference is expected from the redshift-dependent detectability of broad emission lines:
at higher redshifts, the lower line luminosities associated with lower-mass IMBHs become 
increasingly difficult to detect at the available spectral S/N, 
resulting in a natural bias toward higher-mass objects.

In the \mbh--$L_\mathrm{\hb}$ plane, 
the $0.5<z<1$ sample extends to substantially higher broad-\hb\ luminosities than the $z<0.6$ sample, 
with $L_{\hb} \sim 10^{41}-10^{42}$\ergs.
The upper envelope of the $L_\mathrm{\hb}$ distribution is also markedly higher than that of the low-redshift sample.
The absence of low-luminosity sources at $0.5<z<1$ is likely affected by the limited sensitivity of the observations, but the presence of IMBH AGNs reaching such high broad-line luminosities is unlikely to be solely a selection effect.
Consistent with this higher luminosity range,
the \mbh--\redd\ distribution shows that the present sample contains a substantial population of highly accreting IMBH AGNs, 
with \redd$\sim1.2-9.0$,
and generally occupies higher Eddington ratios than the $z<0.6$ comparison sample.

This behavior is consistent with and extends to higher redshift the evolutionary trend reported by \citet{imbh_liu26},
who found a systematic decline in the upper envelopes of both the Eddington ratio and broad-\ha\ luminosity distributions for IMBH AGNs over $0<z\lesssim0.6$.
In that study, the evolution was characterized using the maximum-luminosity envelope in successive redshift bins, a quantity that is relatively insensitive to Malmquist bias in flux-limited samples.
The higher luminosities and Eddington ratios reached by the present $0.5<z<1$ sample therefore extend this trend into the intermediate-redshift regime. 
The high Eddington ratios of the present sample also provide a useful context for the emission-line properties examined in the following sections.

\subsection{IMBH AGNs on the Optical EV1 Sequence}

\begin{figure}[htbp]
   \centering
   \includegraphics[width=0.495\textwidth]{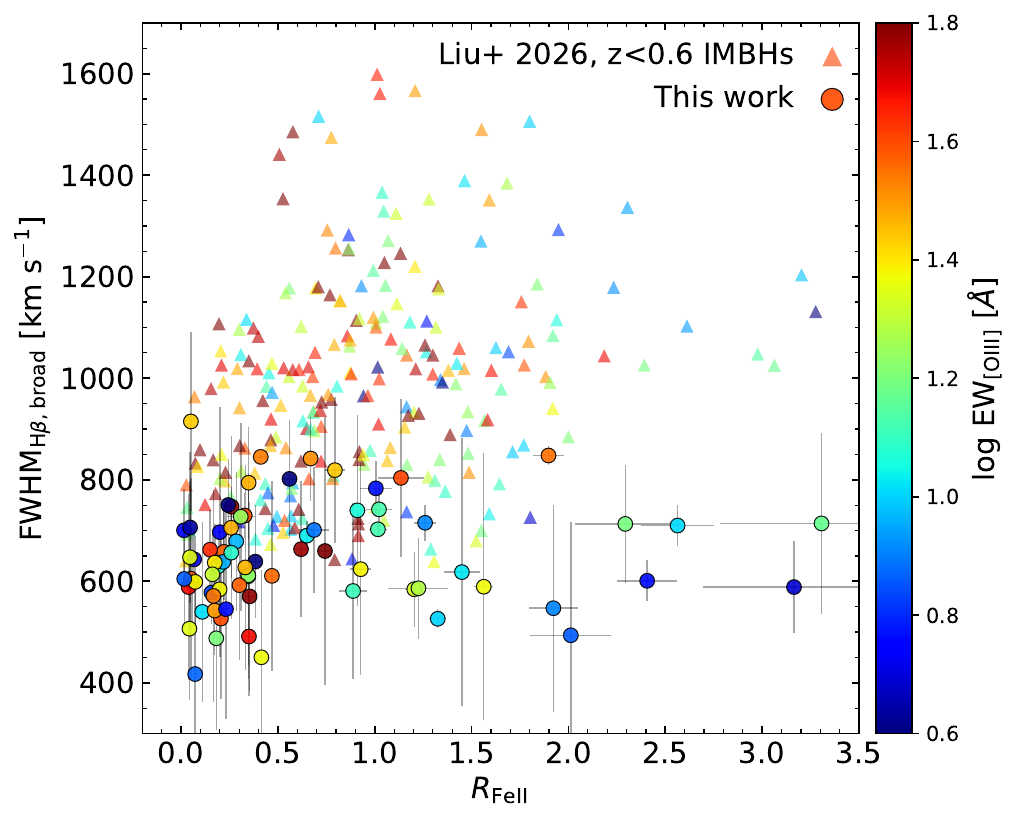}
   \caption{Location of the IMBH AGNs in the parameter space associated with the optical EV1 sequence,
   defined by the relative strength of the optical \feii\ emission, 
   $R_{\feii} \equiv F_{\feii\,\lambda4570}/F_{\hb^\mathrm{b}}$,
   and the FWHM of the broad \hb\ component.
   Filled circles denote the IMBH AGNs at $0.5<z<1$ identified in this work,
   and triangles denote the IMBH AGNs at $z<0.6$ from \citet{imbh_liu26}, 
   selected with the same \hb-based mass calibration (\S\ref{subsec:distributions}).
   Symbols are color-coded by the rest-frame equivalent width of \oiii$\lambda5007$, log$EW_{\oiii}$.}\label{fig:fig6_EV1}
\end{figure}

The optical Eigenvector 1 (EV1) sequence, 
originally identified in optical spectra of quasars,
describes the systematic correlations among several emission-line properties,
most prominently the decrease of \oiii\ strength toward stronger optical \feii\ strength \citep[e.g.,][]{bg92, shen_ho2014}.
The EV1 sequence is commonly associated with variations in the accretion state,
with higher Eddington-ratio sources generally exhibiting stronger \feii\ emission and weaker \oiii\ emission.
While this framework has been extensively established for luminous quasars,
whether IMBH AGNs follow the same emission-line sequence remains largely unexplored.

Figure~\ref{fig:fig6_EV1} presents the distribution of our IMBH AGNs in the parameter space associated with the EV1 sequence,
defined by the broad \hb\ FWHM and the relative optical \feii\ strength $R_{\feii} (\equiv EW_{\feii\,\lambda4434-4684}/EW_{\hb^{\rm b}}$).
The $0.5<z<1$ IMBH AGNs identified in this work are shown together with the previously published $z<0.6$ IMBH sample
\citep{imbh_liu26}.
The IMBH AGNs occupy the narrower broad-line-width regime than luminous quasars,
while the two IMBH samples span a broadly similar range of $R_{\feii}$,
broadly overlapping the range occupied by quasars.

The color-coded \oiii\ equivalent widths show that IMBH AGNs broadly follow the expected EV1 trend:
sources with stronger \feii\ emission generally exhibit weaker \oiii\ emission,
consistent with the quasar EV1 sequence reported by previous studies \citep[e.g.,][]{bg92,Kzhang2011,shen_ho2014}.
However, substantial scatter exists around this general trend.
Several objects exhibit relatively strong \feii\ emission while retaining moderate or large \oiii\ EWs.
To explore the origin of these outliers, 
we further inspect sources located at the high-\oiii\ EW and high-\feii\ end of the EV1 parameter space.
Among the seven objects with $EW_\mathrm{\oiii}>10^{1.4}$\,\AA\ and $R_{\feii}>0.5$,
five show prominent blueshifted \oiii\ wing components,
with the wing component contributing more than 70\% of the total \oiii\ flux.
This suggests that part of the scatter around the global EV1 relation may be associated with variations 
in the \oiii\ kinematic components.
Such a possibility is consistent with previous quasar studies showing that the \oiii\ core component follows the EV1 trends more closely,
whereas the blueshifted wing component exhibits a weaker dependence on EV1-related parameters
\citep{Kzhang2011,shen_ho2014}.

Overall, the IMBH AGNs broadly follow the EV1 sequence established for more massive quasars,
with their global optical emission-line properties consistent with known accretion-related trends;
However, the \oiii\ profile carries additional kinematic information beyond its integrated strength.
The relatively weak connection between the \oiii\ wing component and EV1-related parameters reported in previous studies
\citep[e.g.,][]{Kzhang2011,shen_ho2014} suggests that the detailed \oiii\ profile shape may provide complementary information 
on the narrow-line region beyond the instantaneous accretion state.
We therefore investigate the \oiii\ kinematics of the IMBH AGNs in the following section.

\subsection{\oiii\ Profiles: Broader with Stronger Blueshifted Wings \label{subsec:o3profile}}

During the spectral decomposition, 
we find that the \oiii$\lambda5007$ profiles in the $0.5<z<1$ IMBH AGNs often appear broader and more asymmetric than those in the lower-redshift comparison sample.
In individual fits, an additional blueshifted component is frequently required to reproduce these broader profiles.
We therefore further compare the \oiii\ kinematic properties of the two samples.

\begin{figure}[htbp]
   \centering
   \includegraphics[width=0.46\textwidth]{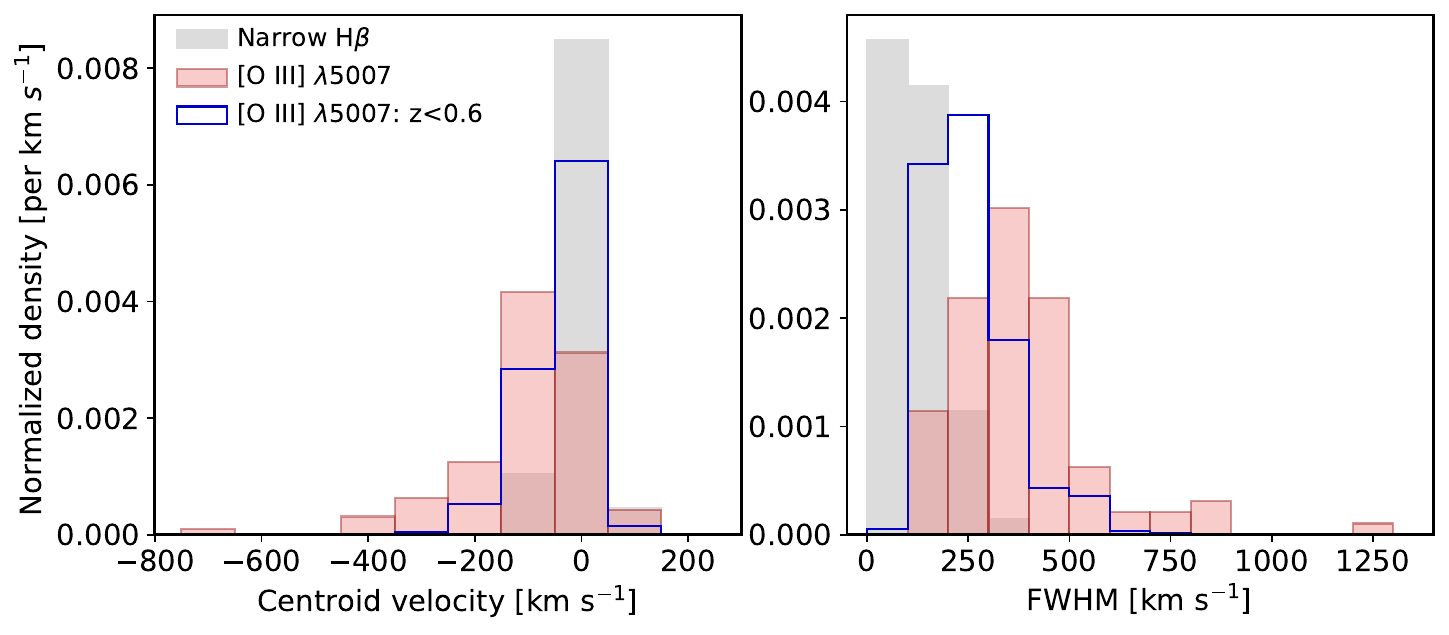}
   \caption{Normalized distributions of the \oiii$\lambda5007$ centroid velocity and FWHM for the $0.5<z<1$ IMBH AGN sample.
   Left panel: centroid velocity distribution for \oiii$\lambda5007$; Right panel: corresponding FWHM distribution. 
   The red filled histograms show the results for the current sample, 
   while the blue step histograms show the $z<0.6$ comparison sample \citep{imbh_liu26}.
   Because the two samples contain different numbers of objects, 
   the histograms are displayed as normalized densities such that the total area in each histogram integrates to unity within each sample. 
   For reference, the gray filled histograms show the corresponding distributions of the narrow \hb\ measured for the $0.5<z<1$ sample.
   }\label{fig:fig7_o3distr}
\end{figure}

Figure \ref{fig:fig7_o3distr} shows the normalized distributions of the \oiii$\lambda5007$ 
centroid velocity and FWHM for the $0.5<z<1$ IMBH AGN sample and the $z<0.6$ comparison sample from \citet{imbh_liu26}. 
As the latter sample is substantially larger, we show normalized distributions, 
with each histogram scaled to unit total area, rather than raw counts.
The narrow \hb\ distribution of the $0.5<z<1$ sample is also shown as a reference.

The \oiii\ centroid velocity distribution shows a substantial blueshift relative to the narrow \hb\ reference,
while the \oiii\ FWHM distribution is markedly broader.
More interestingly, the \oiii\ distributions reveal systematic differences between the two redshift ranges.
The $z<0.6$ sample has centroid velocities predominantly close to the systemic velocity,
with some evidence for blueshifts,
whereas the $0.5<z<1$ sample shows a substantially higher incidence of blueshifted \oiii\ profiles.
The FWHM distribution of the higher-redshift sample is likewise shifted toward broader values than that of the $z<0.6$ sample.
Overall, the \oiii\ profiles differ systematically between the two redshift ranges, with the $0.5<z<1$ sample showing more pronounced blueshifts and broader line widths.

\begin{figure}[htbp]
   \centering
   \includegraphics[width=0.49\textwidth]{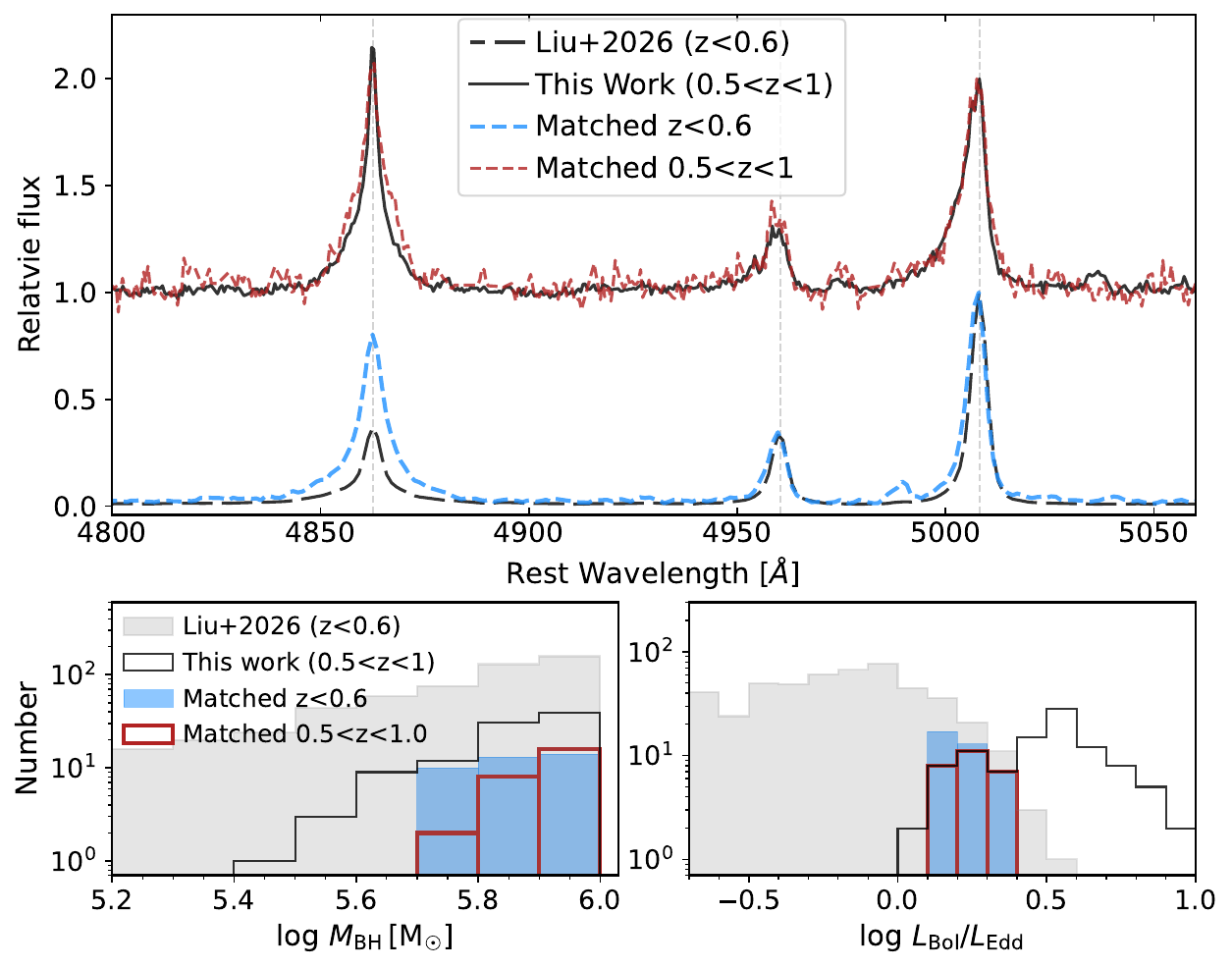}
   \caption{Top panel: Composite \hb\ and \oiii$\lambda5007$ spectra for the $0.5<z<1$ and $z<0.6$ IMBH AGN samples.
   The solid ($0.5<z<1$) and long-dashed ($z<0.6$) black lines show the full sample composite construed from all objects in each sample.
   The red and blue dashed lines indicate the matched subsample composites restricted to sources with overlapping BH mass 
   and Eddington ratio distributions.
   All spectra are normalized to the \oiii\ peak flux and with no smoothing.
   Bottom panels: BH mass (left) and Eddington ratio (right) distributions of the matched subsamples for 
   $0.5<z<1$ (red open bars) and $z<0.6$ (blue filled bars) IMBH AGN samples.
   For reference, the gray shaded step histograms and black open step histograms show the corresponding distributions of the 
   full $0.5<z<1$ and $z<0.6$ samples, respectively.
   The matched subsamples are restricted to a narrow range of \mbh=$10^{5.7}-10^{6.0}$ \msun\ and \redd=$10^{0.1}-10^{0.4}$.
   This restricted parameter range was chosen such that the BH mass and Eddington ratio distributions of 
   the two matched subsamples are as similar as possible, 
   enabling a fair comparison of \oiii\ and \hb\ properties independent of sample selection effects driven by these physical parameters.
   }\label{fig:fig8_o3composite}
\end{figure}

Figure~\ref{fig:fig8_o3composite} further illustrates the differences between the two redshift samples through composite spectra.
Composite spectra reduce object-to-object variations and highlight systematic differences in the average \hb\ and \oiii\ profiles.
We construct the composites by equally weighting the individual spectra and normalizing each spectrum to the peak flux of \oiii$\lambda5007$.
For the full-sample composites, we combine 98 sources from the $0.5<z<1$ sample and 558 sources from the $z<0.6$ comparison sample with reliable broad-\hb\ detections.

The full-sample composites show clear differences in the \oiii\ profile.
Compared with the $z<0.6$ composite, the $0.5<z<1$ composite exhibits a broader \oiii\ profile with a more pronounced blueshifted wing and stronger blue asymmetry, consistent with the distributions shown in Figure~\ref{fig:fig7_o3distr}.
The broader and more asymmetric \oiii\ profile is unlikely to result from differences in instrumental spectral resolution, since the DESI spectra used for the $0.5<z<1$ sample have higher spectral resolution than the SDSS spectra used for the $z<0.6$ sample.

The two composites also show a clear difference in the integrated \oiii/\hb$_\mathrm{total}$ flux ratio, 
where the \hb\ flux includes both broad and narrow components.
This ratio is related to the EV1 sequence, 
along which the strengths of \feii\ and \oiii\ emission, 
both measured relative to broad \hb, are anti-correlated \citep{bg92};
thus, \oiii/\hb$_\mathrm{broad}$ can serves as a rough tracer of position along this sequence.
Rather than performing a broad/narrow decomposition to isolate broad \hb, 
we use the total \hb\ flux, since the relative strengths of the \oiii\ and \hb\ emission are readily apparent from the composite spectra themselves.
The $0.5<z<1$ composite shows an apparently lower \oiii/\hb\ ratio than the $z<0.6$ composite,
which is broadly consistent with the different Eddington-ratio and luminosity distributions of the two samples.
Higher-Eddington-ratio AGNs tend to have weaker \oiii\ emission relative to broad \hb\ along the EV1 sequence, 
while the \oiii\ equivalent width decreases with increasing luminosity due to the Baldwin effect \citep[e.g.,][]{kzhang2013,shen_ho2014}.


To examine whether the differences in \oiii\ profiles are driven by the different distributions 
of fundamental AGN parameters between the two redshift samples, 
we construct matched subsamples with similar BH mass and Eddington-ratio distributions 
(Figure~\ref{fig:fig8_o3composite}, bottom panels).
The matched samples contain 26 objects at $0.5<z<1$ and 37 objects at $z<0.6$, respectively.
Because the matched samples occupy similar ranges in BH mass and Eddington ratio, 
they also mitigate the systematic luminosity differences expected from the virial mass relation and flux-limited selection.
We compare their composite spectra in Figure~\ref{fig:fig8_o3composite} (top panel, red and blue dashed curves).

Despite the close match in BH mass and Eddington ratio, 
the two matched subsamples show clearly different \oiii\ profiles.
The $0.5<z<1$ matched composite retains a broader profile, stronger blue asymmetry, 
and more extended blueshifted wings than the matched $z<0.6$ composite.
Thus, the enhanced \oiii\ kinematic signatures seen in the higher-redshift sample persist after controlling for the distributions of BH mass and Eddington ratio.

We further examine the individual-object \oiii\ kinematics of the matched subsamples
through both the \oiii\ FWHM and the asymmetric-component decomposition.
Given the relatively small matched sample sizes, 
the difference in the \oiii\ FWHM distribution is statistically suggestive (KS statistic of $D=0.349$ with $p=3.507\times10^{-2}$),
consistent with the broader composite profile from higher-redshift sample.
The asymmetric-component decomposition provides additional information on the origin of this difference.
Among the 26 objects in the $0.5<z<1$ matched sample, 
11 require an additional blueshifted component with a velocity offset exceeding 100 km s$^{-1}$, 
compared with 8 of 37 objects in the matched $z<0.6$ sample.
Although the occurrence rates are comparable, 
the additional component contributes a substantially larger fraction of the total \oiii\ flux in the higher-redshift sample, with mean flux fractions of 0.82 and 0.48, respectively.
The corresponding mean velocity offsets are similar ($-220$ and $-176$ km s$^{-1}$), 
indicating that the primary difference is not a substantially larger velocity shift, 
but rather the stronger contribution of the asymmetric component.

Interestingly, the matched composites closely resemble their corresponding full-sample composites, 
despite the restricted parameter ranges of the matched samples.
The matched $z<0.6$ subsample is drawn primarily from the high-Eddington-ratio end of the full lower-redshift sample, 
whereas the matched $0.5<z<1$ subsample occupies a relatively lower-Eddington-ratio regime within the higher-redshift population.
The persistence of similar profile differences across these different accretion regimes suggests that 
the strength of the \oiii\ wing is not strongly controlled by the instantaneous Eddington ratio.
This behavior is consistent with previous studies of more luminous AGNs,
which found only a weak dependence of the \oiii\ wing component on Eddington ratio \citep[e.g.,][]{Kzhang2011,shen_ho2014}.

We also examine the \oiii/\hb$_\mathrm{total}$ flux ratio in the matched subsamples.
The flux ratio of the matched $0.5<z<1$ subsample is close to that of the corresponding full sample, 
whereas that of the matched $z<0.6$ subsample is lower than its full-sample value.
This difference is broadly consistent with their locations within the parent Eddington-ratio distributions: 
the $z<0.6$ matched subsample is drawn from the high-Eddington-ratio end of a relatively broad distribution, whereas the $0.5<z<1$ matched subsample, although toward the lower-Eddington-ratio end, remains close to the dominant range of its parent sample.
The two matched samples nevertheless show different \oiii/\hb\ ratios despite having similar Eddington-ratio distributions, 
possibly reflecting the substantial scatter in the line ratio, differences in the relative contributions of the broad and narrow \hb\ components, and the stronger non-core \oiii\ contribution in the $0.5<z<1$ sample.
Overall, the behavior of \oiii/\hb\ is broadly consistent with its dependence on Eddington ratio, 
while the residual difference between the matched samples suggests that the relation is not one-to-one.

Together, the matched comparison demonstrates that the enhanced \oiii\ kinematic signatures in the $0.5<z<1$ sample persist after controlling for BH mass and Eddington ratio，while the \oiii/\hb\ ratio exhibits a trend broadly consistent with the Eddington-ratio distributions of the two samples.

Since the enhanced blueshifted \oiii\ wing component may be associated with outflowing ionized gas,
it is natural to ask whether similar kinematic signatures are present in other high-ionization forbidden lines,
such as \neiii$\lambda3869$ and \nev$\lambda3427$.
However, these lines are substantially weaker than \oiii$\lambda5007$ in our $0.5<z<1$ IMBH sample and generally lack sufficient S/N for reliable profile decomposition.
A systematic comparison of their kinematic components will therefore require higher-quality spectroscopy.

\section{Discussion}

\subsection{Probable Evidence for Evolution of IMBHs AGNs at $z<1$}\label{subsec:evolution}

The comparison between the $0.5<z<1$ and $z<0.6$ IMBH AGN samples reveals significant differences 
in both their Eddington-ratio distributions and \oiii\ kinematics,
providing two complementary observational perspectives on the evolution of IMBH growth. 

In our previous study of $z<0.6$ IMBH AGNs \citep{imbh_liu26}, 
we found that the upper envelopes of both the Eddington ratio and 
broad \ha\ luminosity distributions systematically decline toward lower redshift.
This trend was quantified using the maximum-luminosity envelope in successive redshift bins, 
which is relatively insensitive to Malmquist bias in flux-limited samples.
The present $0.5<z<1$ sample reaches substantially higher broad-\hb\ luminosities 
and Eddington ratios than those typically observed at $z<0.6$,
extending this previously identified trend to earlier cosmic epochs.
In particular, the higher upper envelope of the Eddington-ratio distribution 
indicates that IMBHs at earlier epochs could reach more extreme accretion states 
than their lower-redshift counterparts.
The combined results from the two redshift ranges therefore suggest 
an evolution in the conditions under which IMBHs can reach their most extreme accretion states.

A second notable difference is found in the \oiii\ emission-line profiles.
The stronger blueshifted wing emission in the $0.5<z<1$ sample shows 
an almost independent behavior with respect to Eddington ratio in our matched comparison.
This is consistent with previous studies of more luminous AGNs,
which found that the \oiii\ wing component has only a weak dependence on 
accretion-related EV1 parameters \citep[e.g.,][]{Kzhang2011,shen_ho2014}.
Previous work has suggested that the \oiii\ wing is mainly influenced by factors 
such as the orientation of the system \citep{shen_ho2014} 
or the physical conditions of the gas in the host galaxy or narrow-line region \citep{Kzhang2011}.
The systematic difference in wing strength between our two redshift ranges therefore raises 
the possibility that the ionized-gas environments surrounding actively growing IMBHs evolve with cosmic time.
In particular, changes in the amount, spatial distribution, 
or kinematic state of the gas available around the nucleus may affect the strength and profile of the outflowing \oiii\ component.
Interestingly, recent JWST observations of luminous quasars at $z\sim5-6$ have found a 
substantially higher incidence of powerful galaxy-scale \oiii\ outflows than in lower-redshift comparison samples \citep{o3_liuwz26}.
Although these quasars are much more massive and luminous than the IMBH AGNs studied here,
their enhanced ionized-gas activity at earlier cosmic epochs provides 
a higher-redshift analogue to the stronger \oiii\ wing observed in our $0.5<z<1$ sample.
Together, these observations raise the possibility that the gas environments surrounding 
actively accreting BHs evolve over cosmic time.

Overall, the $0.5<z<1$ sample significantly differs from the lower-redshift IMBH population 
in both the maximum accretion rates reached and the strength of the \oiii\ wing emission. 
The former directly traces the evolution of the most extreme accretion states, 
while the latter suggests that changes in the surrounding ionized-gas environment 
may provide an additional signature of the evolving conditions under which IMBHs grow.

\subsection{Implication for BH Seeds and Rapid Growth at $z<1$}

The IMBH AGNs identified at $0.5<z\lesssim1$ occupy a BH mass range 
that is closely connected to the BH seed population.
In this mass regime, episodes of rapid accretion represent an important phase 
of BH mass assembly and are directly relevant to the growth of seed-mass BHs.
The ability of IMBHs to reach high and, in some cases, 
super-Eddington accretion rates at these relatively late cosmic epochs therefore demonstrates 
that rapid growth of seed-mass BHs is not restricted to the earliest stages of cosmic history.

At much earlier cosmic time, 
JWST observations have uncovered numerous low-mass BHs at $z>4$ accreting 
near or above the Eddington limit \citep[e.g.,][]{GNZ11_2024,Suh2025,Baccus2025}. 
These systems are commonly invoked in discussions in the context of the formation and rapid growth of the earliest SMBHs.
The rapidly growing IMBHs identified here extend this picture to substantially later cosmic epochs:
BHs with masses close to the seed regime can still undergo episodes of rapid growth at $z<1$.
The presence of rapidly growing, seed-mass BHs at these relatively late epochs raises the question 
of whether they represent the continued growth of early-formed seeds or a younger population of BH seeds formed at later times.
In the former scenario, early-formed seeds could have remained largely quiescent 
before being re-activated through suitable fueling processes \citep{Natarajan2021}. 
In the latter, additional seed-formation channels may operate at lower redshifts, 
such as runaway stellar collisions in dense stellar clusters \citep[e.g.,][]{Gurkan2004,Fragione2022,Haberle2024,Huang2025}. 
The present observations do not distinguish between these possibilities, 
but they demonstrate that efficient growth of seed-mass BHs can 
continue well after the epoch of the earliest massive BHs.

The declining upper envelope of the Eddington-ratio distribution toward lower redshift 
further suggests that the conditions required to reach the most extreme accretion states 
become progressively less accessible with cosmic time.
A quantitative assessment of the evolution in the fraction of super-Eddington accretors 
requires a complete Eddington-ratio distribution function and is beyond the scope of this work.
Nevertheless, the evolution of the upper envelope may reflect changes in the availability 
of cold gas or in the efficiency of mechanisms capable of transporting gas to the nuclear region.
Understanding these fueling conditions will therefore be important for determining 
why rapid growth of seed-mass BHs can persist to $z<1$,
but becomes increasingly difficult to sustain toward the present day.

A natural next step is to investigate the host galaxies of intermediate-redshift IMBH AGNs.
Such observations could test whether rapidly growing IMBHs reside in distinctive gas-rich, compact, 
or dynamically disturbed environments, 
and whether their host properties connect them to the low-mass BHs and ``little red dots'' 
revealed by JWST at much earlier cosmic epochs \citep[e.g.,][]{Akins2025,chenchanghao2025,Inayoshi2025}.


Further insights come from comparison with local IMBH AGN samples.
As discussed in \S~\ref{subsec:evolution}, 
the maximum Eddington ratios attained by IMBH AGNs decline toward lower redshift.
While a quantitative assessment of changes in the fraction of super-Eddington accretors requires 
construction of the Eddington ratio distribution function and is beyond the scope of this work, 
the observed decrease in the upper envelope already points to an evolution in the conditions required to trigger and sustain high accretion. 
This trend may reflect a progressive reduction in cold gas availability or in the efficiency of mechanisms capable of driving gas to the nuclear region, 
with direct consequences for the late-time activation of BH seeds.

Finally, recent JWST observations have revealed a population of compact, red objects at $z > 4$, 
often referred to as ``little red dots'' (LRDs).
These objects typically host BHs with masses $10^{6}-10^{8}$\msun\
that appear over-massive relative to expectations from the local \mbh-\mgal\ relation, 
indicating that rapid BH growth can occur prior to, 
or at least outpace, galaxy assembly at early epochs \citep[e.g.,][]{Akins2025,chenchanghao2025,Inayoshi2025}.
In the local Universe, observations of dwarf galaxies have uncovered a growing number of IMBH 
candidates that are displaced from galaxy centers \citep[e.g.,][]{Reines2020,Mezcua2020,Sturm2026}. 
Such off-nuclear IMBHs have been interpreted as BHs delivered by minor mergers of satellite galaxies,
gravitational recoil displacements following BH mergers, 
or BHs embedded in dense stellar systems such as nuclear star clusters \citep{Greene2020},
highlighting the complex dynamical and environmental pathways of BH seeds.
Within this diverse landscape, 
the rapidly accreting IMBH AGNs at $0.5 <z \lesssim 1$ represent a clearly 
identifiable phase of intense accretion,
providing an important observational link between high-redshift seed growth and 
the more quiescent IMBH population observed in the local Universe.

\section{Summary}

In this study, we have conducted a systematic search for IMBH AGNs at $0.5 < z \lesssim 1$ using DESI DR1 spectroscopy.
By identifying broad \hb\ emission through quantitative fitting criteria and multi-line consistency checks,
we construct a final sample of 98 broad-line IMBH AGNs with BH masses $\mbh < 10^{6}$\msun.
This sample provides the first systematic census of IMBH AGNs at $0.5<z<1$.
Our main results are summarized as follows:

(1) We identify 98 IMBH AGNs at $0.5<z\lesssim1$

The DESI-selected IMBH AGNs span a BH mass range of $10^{5.5}-10^{6}$ \msun,
broad \hb\ luminosities of $10^{41}-10^{42}$\,\ergs, 
bolometric luminosities of $10^{44}-10^{45}$\,\ergs,
and Eddington ratios from 1.2 to 9.0.
The broad-line identifications are further supported by comparisons with SDSS spectra for nine objects observed by both surveys, 
demonstrating the reliability of the DESI-based measurements and sample selection.

(2) IMBH AGNs Follow the Quasar EV1 Sequence

The IMBH AGNs broadly follow the established quasar EV1 sequence, 
with stronger optical \feii\ emission generally associated with weaker \oiii$\lambda5007$ emission. 
This demonstrates that the overall optical emission-line properties of IMBH AGNs are broadly 
consistent with the accretion-related trends established for more massive AGNs.

(3) $0.5<z<1$ IMBH AGNs Have Broader \oiii\ Profiles and Stronger Blueshifted Wings.

The $0.5<z<1$ IMBH AGNs exhibit broader \oiii\ profiles and a stronger contribution 
from blueshifted wing components than the $z<0.6$ sample. 
These differences persist after matching the two samples in BH mass and Eddington ratio,
while the \oiii\ wing strength in the matched samples is nearly independent of Eddington ratio. 
The enhanced \oiii\ kinematic signatures therefore provide an additional observational distinction 
between the $0.5<z<1$ and $z<0.6$ IMBH populations.

(4) IMBH Growth Conditions Evolve over Cosmic Time.

Compared with the $z<0.6$ population, 
the $0.5<z<1$ sample reaches substantially higher broad-line luminosities and Eddington ratios. 
Together with the stronger \oiii\ kinematic signatures, 
these results indicate that the conditions surrounding actively growing IMBHs differ between the two redshift regimes. 
The evolution of the upper envelope of the Eddington-ratio distribution further suggests that 
the conditions required for IMBHs to reach their most extreme accretion states become progressively less accessible toward lower redshift. 
Because the BH masses in the present sample are close to the expected seed-mass regime, 
the existence of super-Eddington IMBHs with Eddington ratios well above unity at $z<1$ demonstrates that
seed-mass BHs can continue to undergo rapid growth at relatively late cosmic epochs. 
This provides a complementary view of BH seed growth to the rapidly accreting low-mass BHs 
recently identified at much higher redshifts with JWST.

Overall, this work bridges the gap between local IMBH AGN samples and the emerging population 
of rapidly accreting low-mass BHs at high redshift. 
By showing that IMBH AGNs reach more extreme accretion states and 
exhibit systematically different \oiii\ kinematics at higher redshift,
this work provides new observational insights into the cosmic evolutionary pathways of IMBHs,
the late-time growth of seed-mass BHs, 
and the connection between BH growth, ionized-gas outflows, and their host galaxies.


\begin{acknowledgments}
This work is supported by the National Natural Science Foundation of China (12373013), 
the Yunnan Provincial Basic Research Program (202501AT070029), 
and the Open Research Fund of Key Laboratory for Polar Science, Ministry of Natural Resources (KP202405).
L.C.H. was supported by the China Manned Space Program (CMS-CSST-2025-A09), 
the National Science Foundation of China (12233001). 
This work is based on observations obtained by the Dark Energy Spectroscopic Instrument (DESI); 
we acknowledge the entire DESI team for providing the data that made this work possible. 
We also make use of optical spectroscopic data from SDSS, ultraviolet photometry from GALEX, 
and WISE $W1$ and $W2$ photometry provided in the DESI catalog. 
All the {\it GALEX} data used in this paper can be found in MAST: 
\dataset[10.17909/T9H59D]{https://doi.org/10.17909/T9H59D}.
\end{acknowledgments}

\bibliography{imbh}{}
\bibliographystyle{aasjournalv7}


\end{CJK*}
\end{document}